\documentclass[aps,prb,twocolumn,superscriptaddress,amsmath,amssymb,floatfix,longbibliography]{revtex4-2}

\usepackage{amsmath,amssymb,amsfonts}      
\usepackage{graphicx}                      
\usepackage{dcolumn}                      
\usepackage{bm}                           
\usepackage{hyperref}                     
\usepackage[usenames,dvipsnames]{xcolor}  
\usepackage{booktabs}                     
\usepackage{siunitx}                      
\usepackage{mhchem}                  
\usepackage[T1]{fontenc}                  
\usepackage{multirow}                     
\usepackage{physics}                      
\usepackage{soul}
\usepackage{xcolor}

\soulregister\cite7
\soulregister\ref7
\soulregister\text7
\soulregister\ce7

\hypersetup{
  colorlinks=true,
  linkcolor=BlueViolet,
  citecolor=ForestGreen,
  urlcolor=MidnightBlue
}

\begin{document}

\title{Strongly Quenched Kramers Doublet Magnetism in SmMgAl$_{11}$O$_{19}$}


\author{Sonu Kumar}
\email{sonu.kumar@matfyz.cuni.cz}
\affiliation{Charles University, Faculty of Mathematics and Physics, Department of Condensed Matter Physics, Prague, Czech Republic}
\affiliation{Adam Mickiewicz University, Faculty of Physics and Astronomy, Department of Experimental Physics of Condensed Phase, Poznan, Poland}

\author{Barbora Salajov\'a}
\affiliation{Charles University, Faculty of Mathematics and Physics, Department of Condensed Matter Physics, Prague, Czech Republic}

\author{Andrej Kancko}
\affiliation{Charles University, Faculty of Mathematics and Physics, Department of Condensed Matter Physics, Prague, Czech Republic}

\author{Cinthia A. Corr\^ea}
\affiliation{Institute of Physics of the Czech Academy of Sciences, Na Slovance, Prague, Czech Republic}

\author{Shuvajit Halder}
\affiliation{Charles University, Faculty of Mathematics and Physics, Department of Condensed Matter Physics, Prague, Czech Republic}

\author{Ross H. Colman}
\email{ross.colman@matfyz.cuni.cz}
\affiliation{Charles University, Faculty of Mathematics and Physics, Department of Condensed Matter Physics, Prague, Czech Republic}

\date{\today}


\begin{abstract}
We report magnetic susceptibility, isothermal magnetization, and specific-heat
measurements on the rare-earth hexaaluminate SmMgAl$_{11}$O$_{19}$, where Sm$^{3+}$
realizes a strongly quenched Kramers doublet on a triangular lattice with an
exceptionally weak net exchange scale. The Curie--Weiss analysis yields strongly reduced ground-doublet $g$ factors,
$g_{ab}\simeq 0.65$ and $g_{c}\simeq 0.70$.
 This indicates that the low-temperature response is governed
primarily by single-ion physics, with crystal-field splitting and $J$-multiplet
mixing jointly renormalizing the Sm$^{3+}$ moment, rather than collective exchange.
For $H \parallel c$, the specific heat shows no $\lambda$-type anomaly down to
0.35~K but evolves into a well-defined two-level Schottky peak whose gap grows
linearly with field, yielding $g_c\simeq0.62$ and recovering nearly all of $R\ln2$ at high fields,
thereby confirming an effective $S_{\mathrm{eff}}=\tfrac12$ Kramers doublet description
for $T\lesssim10$~K.
 Together, these
results establish SmMgAl$_{11}$O$_{19}$ as a weak-exchange, nearly single-ion
triangular Kramers magnet in which frustration produce an anisotropic low-field correlated regime without inducing
long-range order.
\end{abstract}

\maketitle

\section{Introduction}

Frustrated magnetism provides a fertile route to emergent quantum states that cannot be
captured by conventional mean-field ordering paradigms. In frustrated lattices, competing
interactions or lattice geometry prevent spins from simultaneously satisfying all local
constraints, leading to a macroscopic degeneracy of low-energy configurations and strongly
suppressed ordering temperatures. When combined with low spin, strong spin--orbit coupling, or reduced dimensionality,
geometric frustration suppresses conventional magnetic ordering tendencies,
thereby allowing quantum fluctuations to dominate and potentially stabilize
unconventional ground states such as quantum spin liquids (QSLs), spin glasses,
and multipolar or topological phases.
 \cite{Balents2010,Savary2017,Broholm2020,Ramirez2025}. Identifying real
materials that realize these regimes, and disentangling the role of disorder and
single-ion physics, remains a central challenge in modern condensed-matter research.

Among the simplest geometries supporting strong frustration is the triangular-lattice
antiferromagnet (TLAF). The TLAF has long served as a canonical platform for exploring
correlation-driven physics, but it has gained renewed prominence in the last decade due to
reports of QSL-like behavior in rare-earth TLAFs, most notably the triangular Yb-based
insulator YbMgGaO$_4$ \cite{Li2015,Shen2018YbMgGaO4,Li2015GaplessQSL,Li2016MuonYbMgGaO4,Rao2021}. Although subsequent work has emphasized the
importance of structural disorder in that compound, the broader impact was to establish
rare-earth TLAFs as an experimentally accessible family where crystal-electric-field (CEF)
ground--state doublets, strong spin--orbit coupling, and frustration cooperate to produce highly
renormalized ground states \cite{Zhu2017YbMgGaO4,Li2017CEF}. This momentum has driven extensive
searches for QSL and related phenomena across other frustrated lattices, including kagome
systems such as herbertsmithite, kapellasite and Zn-barlowite \cite{Huang2021Herbertsmithite,Norman2016Herbertsmithite,Shaginyan2013Herbertsmithite,Smaha2020Barlowite,Tustain2020ZnBarlowite}, and pyrochlore
magnets hosting spin-ice, quantum spin-ice, and spin-liquid regimes
\cite{Bramwell2001SpinIce,Huang2016SpinIceHeisenberg,Gingras2014QSI,Yao2020PyrochloreU1,An2025NonKramersPyrochlore,Greedan2006Pyrochlores,Savary2016BreathingPyrochlore,Desrochers2022Ce2Zr2O7,Yahne2024Ce2Sn2O7}. 

\begin{figure*}
    \centering
    \includegraphics[width=1\textwidth]{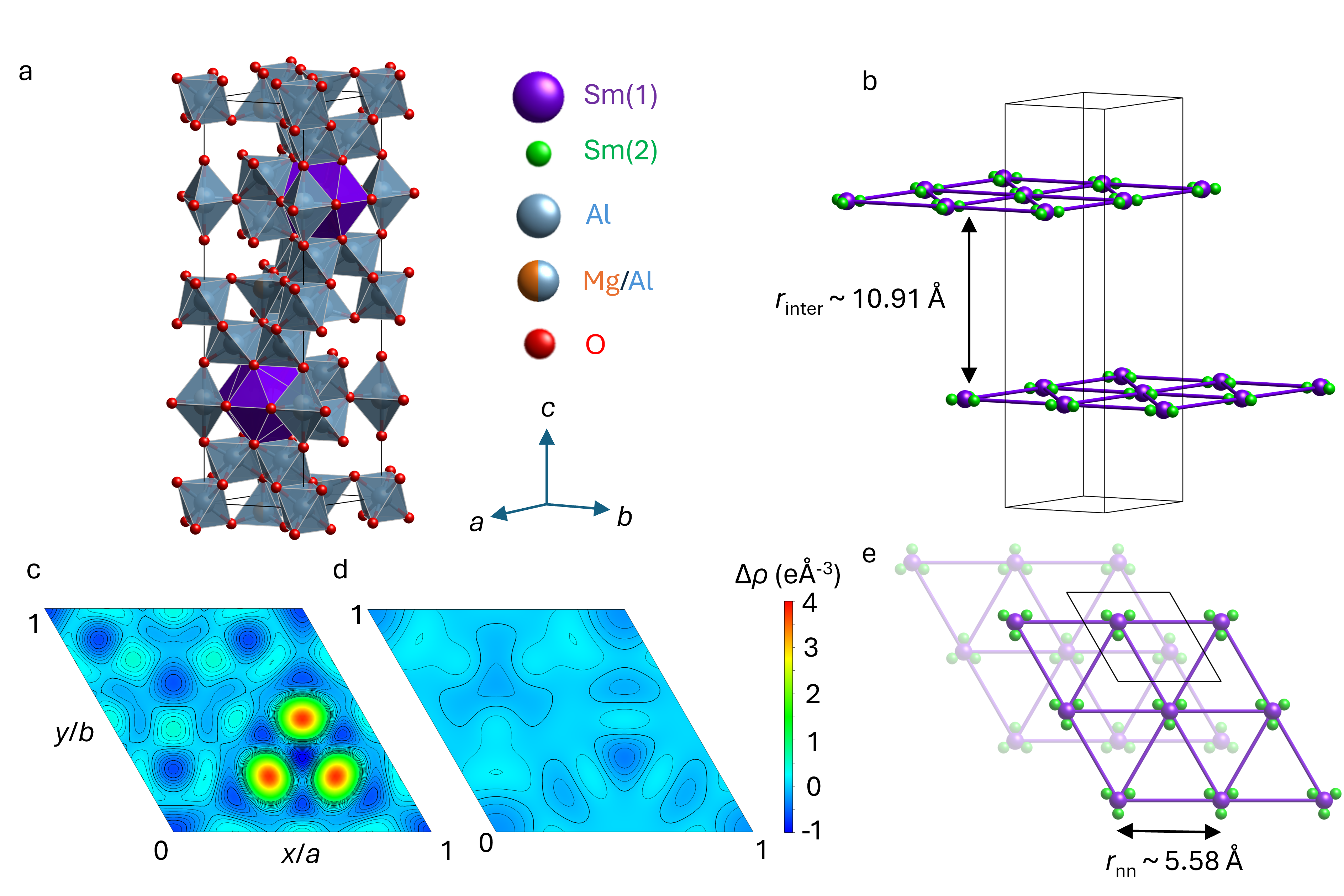}
    \caption{(a) Magnetoplumbite structure of Sm$_{0.941}$MgAl$_{11}$O$_{19}$, with highlighted AlO$_5$ bipyramids and SmO$_{12}$ polyhedra. (b) Triangular layers of Sm$^{3+}$ ions, vertically separated by $r_{\rm inter} = c/2 \approx$ 10.91 $\mathring{\mathrm{A}}$. Fourier difference map projection at $z/c = 0.25$, showing the electronic density difference (c) before and (d) after the inclusion of the off-centered Sm(2) ions on the 6$h$ site. (e) Triangular lattice of Sm(1) and Sm(2) ions projected along the $c$ axis, with the Sm(1)-Sm(1) distance $r_{\rm nn} = 5.58 \mathring{\mathrm{A}}$.}
    \label{fig:Fig1}
\end{figure*}

Rare-earth magnetoplumbite hexaaluminates LnMgAl$_{11}$O$_{19}$ constitute a particularly
attractive but still emerging class of frustrated magnets. They crystallize in a
centrosymmetric hexagonal magnetoplumbite structure, seen in figure~\ref{fig:Fig1}, in which Ln$^{3+}$ ions form
quasi-two-dimensional triangular layers separated by nonmagnetic Al/Mg--O blocks,
so that antiferromagnetic interactions are expected to be strongest within each
Ln layer and comparatively weak couplings acting between the layers, i.e., the system
realizes a set of isolated triangular-lattice antiferromagnetic planes
\cite{Gasperin1984,Saber1981}.
 The rare-earth
layers host strongly spin--orbit-coupled local moments whose low-energy physics is governed
by CEF-split Kramers or non-Kramers doublets due to the anisotropic ligand environments. As a result, this family spans distinct
quantum limits: Kramers ions (e.g., Ce$^{3+}$, Nd$^{3+}$) can realize
exchange-anisotropic effective spin-$\tfrac12$ magnets, whereas non-Kramers ions (e.g.,
Pr$^{3+}$) allow additional multipolar degrees of freedom and
enhanced coupling between spin and lattice \cite{Bastien2025,Kumar2025,Kumar2025NdMgAl11O19}. Recent studies of
Ce- and Pr-based hexaaluminates already point to exotic low-temperature behavior without
long-range order, including strong single-ion anisotropy, pronounced CEF renormalization,
and field-tunable thermodynamic anomalies \cite{Bastien2025,Kumar2025,Cao2024,Li2024PrMgAl11O19,Ma2024PrMgAl11O19,Tu2024PrMgAl11O19,Cao2025CeMgAl11O19,Gao2024CeMgAl11O19}, underscoring the
potential of this platform for frustration-driven quantum magnetism.

Sm-based quantum magnets are, more broadly, well known as promising candidates for
quantum-disordered ground states. The Sm$^{3+}$ ion carries a small free-ion moment, strong
Van Vleck susceptibility, and CEF excitations. The low moment ensures weak exchange interactions, amplifying the importance of quantum
fluctuations of an $S_{\text{eff}}=\frac{1}{2}$ ground--state doublet and rendering magnetic order unusually fragile \cite{Sanders2017KBaREBO3,Sanders2016RE3Sb3Mg2O14, Bairwa2025SmTa7O19}. In SmTa$_{7}$O$_{19}$ the frustrated lattice and these ingredients have been linked to
persistent spin dynamics and QSL-like phenomenology~\cite{Bairwa2025SmTa7O19}.
A closely related scenario has recently been identified in the Nd analogues:
the triangular-lattice hexaaluminate NdMgAl$_{11}$O$_{19}$ exhibits a magnetic
ground state that closely mirrors that of NdTa$_{7}$O$_{19}$, with strong
frustration and the absence of long-range order down to the lowest measured
temperatures~\cite{Kumar2025NdMgAl11O19,Arh2022}. This one-to-one
correspondence between hexaaluminate and tantalate compounds suggests that
SmMgAl$_{11}$O$_{19}$ may realize SmTa$_{7}$O$_{19}$-like correlated or
QSL-like behavior within the magnetoplumbite structure. Despite this
motivation, a detailed single-crystal study of the Sm hexaaluminate
SmMgAl$_{11}$O$_{19}$ has so far been missing, leaving its intrinsic
anisotropy, thermodynamic ground state, and field response largely unexplored.

In this article we present a comprehensive structural, magnetic, and thermodynamic study of single-crystalline SmMgAl$_{11}$O$_{19}$. Using single-crystal X-ray diffraction, magnetometry, and field-dependent specific heat down to sub-Kelvin temperatures, we establish the quenched Kramers-doublet character of Sm$^{3+}$ on a triangular lattice and quantify the extremely weak exchange scale. The combined data reveal a low-field correlated regime without long-range order down to the lowest measured temperatures and a crossover under field to a nearly ideal Zeeman-split doublet response. These results position SmMgAl$_{11}$O$_{19}$ as a key member of the rare-earth hexaaluminate family and a useful platform for studying how frustration, crystal-field excitations, and admixture of the $J = 7/2$ multiplet renormalize local doublet magnetism.

 \begin{figure}
    \centering

    \includegraphics[width=0.48\textwidth]{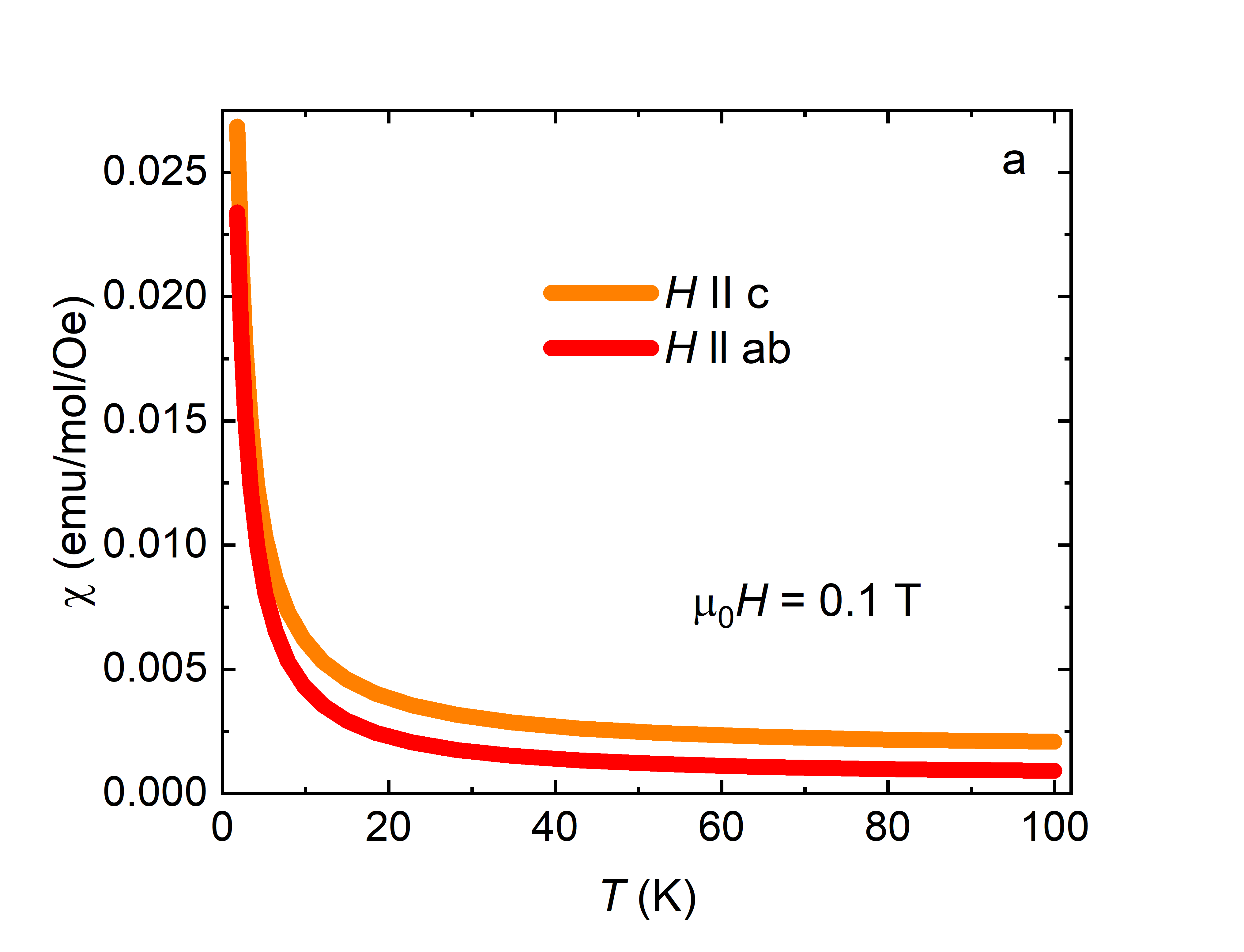}

    \includegraphics[width=0.48\textwidth]{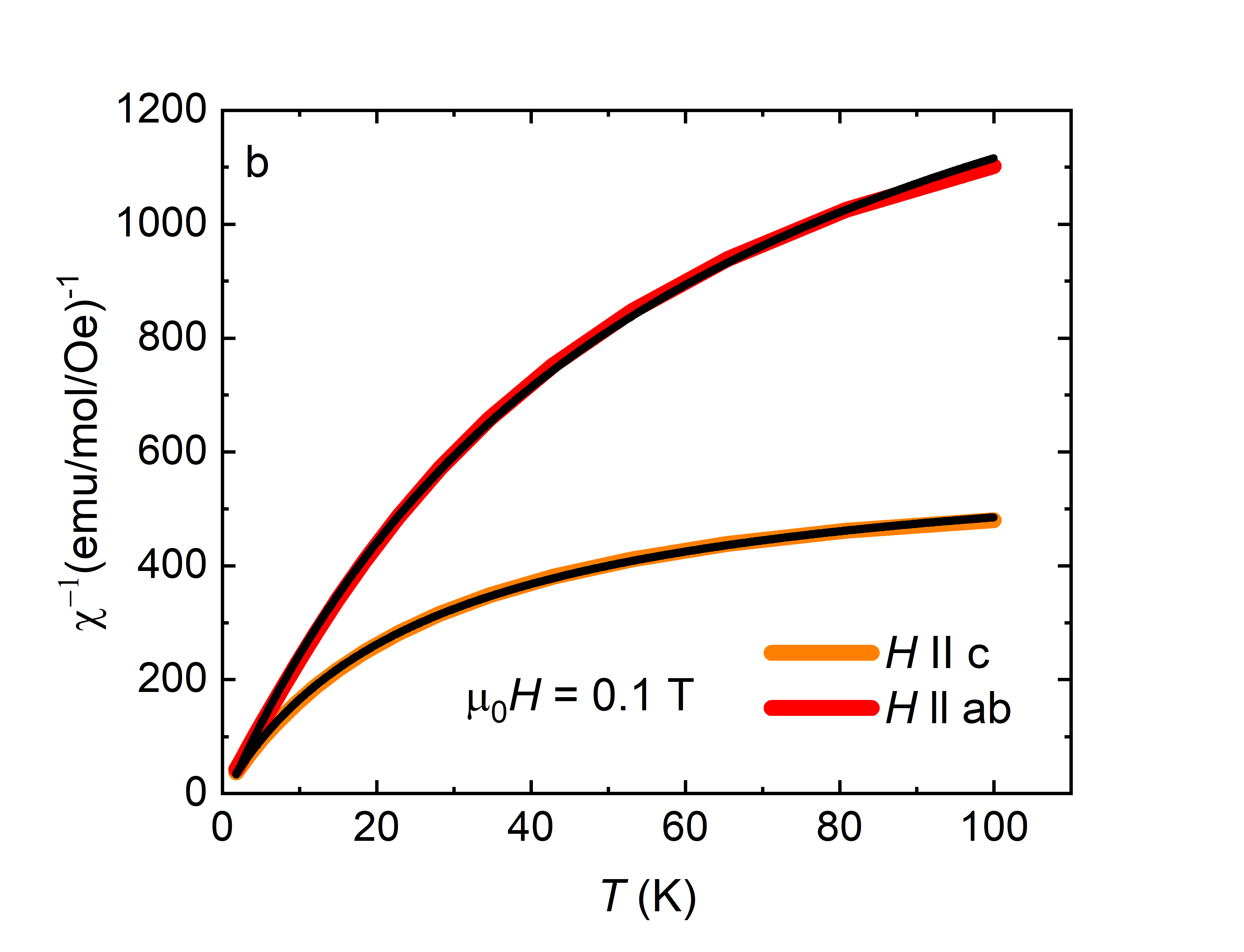}

    \includegraphics[width=0.48\textwidth]{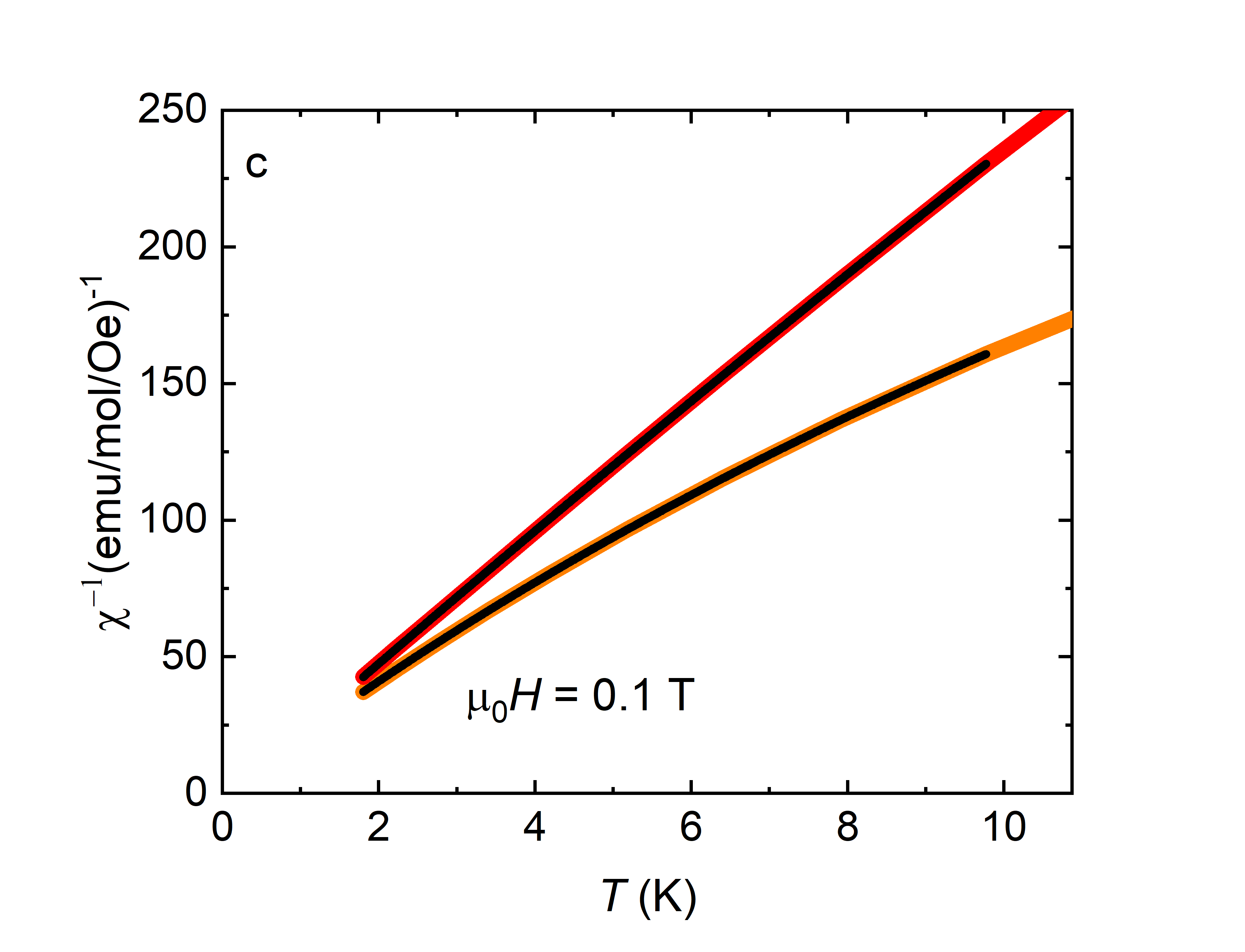}

    \caption{
    (a) DC magnetic susceptibility $\chi(T)$ of SmMgAl$_{11}$O$_{19}$ for 
    $H \parallel c$ and $H \parallel ab$ measured in an applied field of
    $\mu_{0}H = 0.1$~T. 
    (b) Inverse susceptibility $1/\chi(T)$ for both field directions over the
    full temperature range, together with Curie--Weiss fits including a
    Van~Vleck term (solid black lines).
    (c) Low-temperature inverse susceptibility for $H \parallel c$ and 
    $H \parallel ab$ in the 1.8--10~K range, highlighting the linear
    CW+$\chi_{\mathrm{vv}}$ behavior in the $ab$ plane and its absence along $c$.
    Solid black lines show the corresponding CW+$\chi_{\mathrm{vv}}$ fits.
    }
    \label{fig:Fig2}
\end{figure}

\section{Methods}

The synthesis and subsequent crystal growth of SmMgAl$_{11}$O$_{19}$ were
carried out using a conventional solid-state reaction followed by the optical
floating-zone (OFZ) method. Initial precursor binary oxides (Sm$_2$O$_3$, MgO,
and Al$_2$O$_3$, all 99.99\% purity) were calcined at 800~$^\circ$C for 24~h to
remove moisture and carbonate contamination. The oxides were then weighed in
the stoichiometric ratio, thoroughly mixed, and ground to ensure homogeneity of
the mixture. The resulting powder was pressed into cylindrical rods with a
6~mm diameter and a length of about 100~mm. Densification was achieved using a
quasihydrostatic pressure of 2~tons for 15~min, after which the rods were
sintered in air at 1200~$^\circ$C for 72~h to promote the solid-state reaction
and improve the density. 

Single crystals of SmMgAl$_{11}$O$_{19}$ were then grown in a four-mirror OFZ
furnace under an air atmosphere with a slight overpressure of 1~atm and an air
flow rate of 3~L/min. The sintered rods were used both as feed and seed
material. During growth, the feed and seed rods were counter-rotated at
30~rpm to enhance temperature homogeneity and mixing in the molten zone, and
the growth rate was maintained at 2~mm/h.

The single-crystalline nature of the resulting pieces was confirmed using backscattered
Laue X-ray diffraction. A single-crystal X-ray diffraction (SCXRD) experiment was performed at 100\,K on a XtaLAB Synergy R, DW system four-circle diffractometer, using a Mo rotating-anode X-ray tube (Mo K$\alpha$: $\lambda$ = 0.71073\,$\mathring{\mathrm{A}}$) with a mirror monochromator, and a hybrid pixel array detector HyPix-Arc 150$^{\circ}$. Diffraction data were integrated using CrysAlis Pro~\cite{CrysAlisPro2024} with an empirical absorption correction using spherical harmonics~\cite{Clark1995} combined with an analytical numeric absorption correction using a multifaceted crystal model implemented in the SCALE3 ABSPACK scaling algorithm. The structure was solved by charge flipping using the program Superflip~\cite{Palatinus2007} and refined by full-matrix least squares on $F^2$ in Jana2020~\cite{Petricek2023}. 

DC magnetic susceptibility measurements were carried out using a Quantum Design SQUID
Magnetic Property Measurement System (MPMS3). Specific heat was measured from room
temperature down to \SI{0.35}{K} using a Quantum Design Physical Property Measurement System
(PPMS9). Multiple pieces of single crystals were used for the
bulk measurements; the consistency of the results confirms the sample quality and
reproducibility. The data shown here were obtained from crystal pieces with masses of
\SI{7.8}{mg}, \SI{4.72}{mg}, and \SI{1.48}{mg}.

To isolate the magnetic contribution to the specific heat, non-magnetic
LaMgAl$_{11}$O$_{19}$ single crystals were also grown in air by the floating-zone method and
measured under identical conditions. A small low-temperature upturn in the LaMgAl$_{11}$O$_{19}$
specific heat indicates a tiny concentration of magnetic impurities; therefore, in the
low-temperature limit ($T<\SI{1}{K}$) the phononic background used for subtraction was taken
from a $T^{3}$ extrapolation of the measured LaMgAl$_{11}$O$_{19}$ heat capacity. The magnetic
specific heat of SmMgAl$_{11}$O$_{19}$ was then obtained as
$C_{m}(T)=C_{\mathrm{Sm}}(T)-C_{\mathrm{La}}^{\mathrm{ph}}(T)$.

\section{Results}

\subsection{Structural characterization}

The crystal structure of SmMgAl$_{11}$O$_{19}$ obtained by SCXRD at 100 K is hexagonal, space group $P6_3/mmc$ (\#194), $Z$ = 2, and with unit cell parameters $a$ = 5.57895(8) $\mathring{\mathrm{A}}$ and c = 21.8172(3) $\mathring{\mathrm{A}}$, in
line with the expected magnetoplumbite structure. Due to the identical electronic configuration of
Mg$^{2+}$ and Al$^{3+}$ and, therefore, very similar X-ray atomic scattering factors, it is practically impossible to distinguish the site of Mg$^{2+}$ ions site via SCXRD. Therefore, an initial model of SmAl$_{12}$O$_{19}$ was
assumed, giving GOF = 7.56\% and $R_{obs}$ = 6.01\%. Allowing the Sm(1) occupancy to refine led to
a noticable improvement of the model, with occ(Sm1) = 0.902(6), GOF = 6.47\%, and $R_{obs}$ = 5.54\%, indicating a $\sim$10\% Sm$^{3+}$ deficiency. Neutron diffraction experiments on the isostructural CeMgAl$_{11}$O$_{19}$ \cite{Gao2024CeMgAl11O19} suggest that Mg$^{2+}$ is shared with Al$^{3+}$
on the Al(4) site. Splitting the Al(4) site between Mg and Al with fixed 0.5/0.5 occupancies, with
harmonic ADPs and coordinates constrained to be equal, the quality of the refinement practically
stays unchanged - GOF = 6.49\% and $R_{obs}$ = 5.53\%. Individually placing the Mg ion on each of
the remaining Al sites did not lead to an improvement of the model. In the magnetoplumbite
structure, the Al(5) ion is often distributed between two off-centered positions in the oxygen
bipyramid, with the $z$-coordinate refined to $z$(Al5) = 0.2434(4), resulting in the off-centering
distance $\delta$ = 2(0.25 - $z$(Al5))$c$ = 0.29 $\mathring{\mathrm{A}}$. As was found previously, a significant improvement of
the model was reached by adding a Sm(2) atom to the 6$h$ site at a distance of 0.855 $\mathring{\mathrm{A}}$ from Sm(1)
located on the 2$d$ site [8]. Inspection of the Fourier difference maps in the $ab$-plane cut at $z/c$ =
0.25 (figures~\ref{fig:Fig1}c and d) indicates a partially occupied atomic site (density 3.49 $e\mathring{\mathrm{A}}^{-3}$
and charge 1.41 $e$).
Inclusion of this positionally disordered (off-centered) Sm(2) ion results in a substantial
improvement of the agreement factors and yields GOF = 2.63\% and $R_{obs}$ = 2.85\%, and refined occupancies occ(Sm1) = 0.862(2) and occ(Sm2) = 0.0264(8). This solution gives the final formula Sm$_{0.941}$MgAl$_{11}$O$_{19}$, suggesting a $\sim$6\% total Sm$^{3+}$ deficiency. A final improvement to the model was reached by employing an isotropic Becker-Coppens (Type 1, Gaussian) extinction parameter $G_{iso}$ = 0.244(14), leading to the final agreement factors GOF = 2.06\% and $R_{obs}$ = 1.82\%. 


\begin{table*}[t]
\caption{Summary of SCXRD structure solution results.}
\begin{tabular}{cccccccccccccccccccccccccc}
\hline
\multicolumn{4}{c}{\begin{tabular}[c]{@{}c@{}} Chemical composition \\ Sm$_{0.941}$MgAl$_{11}$O$_{19}$ \\ \end{tabular}} & \multicolumn{5}{c}{\begin{tabular}[c]{@{}c@{}}Crystal system, space group \\ hexagonal, $P6_3/mmc$ (\#194) \\ $Z$ = 2 \end{tabular}} & \multicolumn{6}{c}{\begin{tabular}[c]{@{}c@{}}$a$ = 5.57895(8) $\mathring{\mathrm{A}}$ \\ $c$ = 21.8172(3) $\mathring{\mathrm{A}}$ \\ $~~~V$ = 588.078(14)
 $\mathring{\mathrm{A}}^3$\end{tabular}} & \multicolumn{7}{c}{\begin{tabular}[c]{@{}c@{}}Crystal size \\ 155x92x25 $\mu$m$^3$\end{tabular}} & \multicolumn{4}{c}{\begin{tabular}[c]{@{}c@{}}Density (calculated) \\ 4.3295 g.cm$^{-3}$\end{tabular}} \\ \hline

\multicolumn{4}{c}{\begin{tabular}[c]{@{}c@{}}X-ray tube: Mo K$\alpha$ \\ ($\lambda$ = 0.71073 $\mathring{\mathrm{A}}$) \\ $T$ = 99.9(2) K\end{tabular}} & \multicolumn{5}{c}{\begin{tabular}[c]{@{}c@{}}No. of reflections \\ measured/independent/observed \\ 20883/695/638\\ Parameters refined: 49\end{tabular}} & \multicolumn{6}{c}{\begin{tabular}[c]{@{}c@{}}Final $R$ indices \\ $R_{obs}$ = 1.82\% \\ $wR_2$ = 5.69\% \\ GOF = 2.0634 \end{tabular}} & \multicolumn{7}{c}{\begin{tabular}[c]{@{}c@{}}$\Delta \rho_{max}$, $\Delta \rho_{min}$ ($e\mathring{\mathrm{A}}^{-3}$) \\ 0.23, $-$0.38\end{tabular}} & \multicolumn{4}{c}{\begin{tabular}[c]{@{}c@{}}Absorption \\ $\mu$ = 5.733 mm$^{-1}$\end{tabular}} \\ \hline

\multirow{2}{*}{Atom} & \multirow{2}{*}{Site} & \multirow{2}{*}{Symm.} & \multicolumn{3}{c}{\multirow{2}{*}{$x$}} & \multicolumn{2}{c}{\multirow{2}{*}{$y$}} & \multirow{2}{*}{$z$} & \multicolumn{2}{c}{\multirow{2}{*}{Occ.}} & \multicolumn{15}{c}{$U_{\mathrm{ij}}$ ($\times$10$^{-4} \mathring{\mathrm{A}}^2$)} \\ \cline{12-26} 
 & & & \multicolumn{3}{c}{} & \multicolumn{2}{c}{} & & \multicolumn{2}{c}{} & 
\multicolumn{2}{c}{$U_{11}$} & \multicolumn{2}{c}{$U_{22}$} & \multicolumn{2}{c}{$U_{33}$} & 
\multicolumn{3}{c}{$U_{12}$} & \multicolumn{2}{c}{$U_{13}$} & \multicolumn{2}{c}{$U_{23}$} & 
\multicolumn{2}{c}{$U_{\mathrm{eq}}$} \\ \hline

Sm(1) & 2$d$ & $-6m2$ & \multicolumn{3}{c}{$\frac{1}{3}$} & \multicolumn{2}{c}{$\frac{2}{3}$} & $\frac{3}{4}$ & \multicolumn{2}{c}{0.862(2)} & 
\multicolumn{2}{c}{108(1)} & \multicolumn{2}{c}{$U_{11}$} & \multicolumn{2}{c}{39(1)} & 
\multicolumn{3}{c}{$\frac{1}{2}U_{11}$} & \multicolumn{2}{c}{0} & \multicolumn{2}{c}{0} & 
\multicolumn{2}{c}{85(1)} \\

Sm(2) & 6$h$ & $mm2$ & \multicolumn{3}{c}{0.7413(6)} & \multicolumn{2}{c}{0.4826(6)} & $\frac{1}{4}$ & \multicolumn{2}{c}{0.0264(8)} & 
\multicolumn{2}{c}{41(15)} & \multicolumn{2}{c}{$U_{11}$} & \multicolumn{2}{c}{19(16)} & 
\multicolumn{3}{c}{-41(13)} & \multicolumn{2}{c}{0} & \multicolumn{2}{c}{0} & 
\multicolumn{2}{c}{61(12)} \\

Al(1) & 2$a$ & $-3m.$ & \multicolumn{3}{c}{0} & \multicolumn{2}{c}{0} & 0 &
\multicolumn{2}{c}{1} & 
\multicolumn{2}{c}{29(3)} & \multicolumn{2}{c}{$U_{11}$} & \multicolumn{2}{c}{37(4)} & 
\multicolumn{3}{c}{$\frac{1}{2}U_{11}$} & \multicolumn{2}{c}{0} & \multicolumn{2}{c}{0} & 
\multicolumn{2}{c}{32(2)} \\

Al(2) & 4$f$ & $3m.$ & \multicolumn{3}{c}{$\frac{1}{3}$} & \multicolumn{2}{c}{$\frac{2}{3}$} & 0.19026(4) & \multicolumn{2}{c}{1} &
\multicolumn{2}{c}{34(2)} & \multicolumn{2}{c}{$U_{11}$} & \multicolumn{2}{c}{32(3)} & 
\multicolumn{3}{c}{$\frac{1}{2}U_{11}$} & \multicolumn{2}{c}{0}  & \multicolumn{2}{c}{0} & 
\multicolumn{2}{c}{33(2)} \\

Al(3) & 12$k$ & $.m.$ & \multicolumn{3}{c}{0.16755(7)} & \multicolumn{2}{c}{0.33509(4)} & 0.60869(2) & \multicolumn{2}{c}{1} &
\multicolumn{2}{c}{36(2)} & \multicolumn{2}{c}{36(2)} & \multicolumn{2}{c}{40(2)} & 
\multicolumn{3}{c}{$\frac{1}{2}U_{11}$} & \multicolumn{2}{c}{1(1)} & \multicolumn{2}{c}{$\frac{1}{2}U_{13}$} & 
\multicolumn{2}{c}{37(2)} \\

Al(4) & 4$f$ & $3m.$ & \multicolumn{3}{c}{$\frac{1}{3}$} & \multicolumn{2}{c}{$\frac{2}{3}$} & 0.02740(4) & \multicolumn{2}{c}{0.5} &
\multicolumn{2}{c}{23(2)} & \multicolumn{2}{c}{$U_{11}$} & \multicolumn{2}{c}{34(4)} & 
\multicolumn{3}{c}{$\frac{1}{2}U_{11}$} & \multicolumn{2}{c}{0} & \multicolumn{2}{c}{0} & 
\multicolumn{2}{c}{26(2)} \\

Mg(1) & 4$f$ & $3m.$ & \multicolumn{3}{c}{$\frac{1}{3}$} & \multicolumn{2}{c}{$\frac{2}{3}$} & 0.02740(4) & \multicolumn{2}{c}{0.5} &
\multicolumn{2}{c}{23(2)} & \multicolumn{2}{c}{$U_{11}$} & \multicolumn{2}{c}{34(4)} & 
\multicolumn{3}{c}{$\frac{1}{2}U_{11}$} & \multicolumn{2}{c}{0} & \multicolumn{2}{c}{0} & 
\multicolumn{2}{c}{26(2)} \\

Al(5) & 4$e$ & $3m.$ & \multicolumn{3}{c}{0} & \multicolumn{2}{c}{0} & 0.2434(4) & \multicolumn{2}{c}{0.5} &
\multicolumn{2}{c}{49(3)} & \multicolumn{2}{c}{$U_{11}$} & \multicolumn{2}{c}{120(30)} & 
\multicolumn{3}{c}{$\frac{1}{2}U_{11}$} & \multicolumn{2}{c}{0} & \multicolumn{2}{c}{0} & 
\multicolumn{2}{c}{72(11)} \\

O(1) & 6$h$ & $mm2$ & \multicolumn{3}{c}{0.18177(14)} & \multicolumn{2}{c}{0.3635(3)} & $\frac{1}{4}$ & \multicolumn{2}{c}{1} & 
\multicolumn{2}{c}{49(6)} & \multicolumn{2}{c}{115(5)} & \multicolumn{2}{c}{49(6)} & 
\multicolumn{3}{c}{$\frac{1}{2}U_{11}$} & \multicolumn{2}{c}{0} & \multicolumn{2}{c}{0} & 
\multicolumn{2}{c}{78(4)} \\

O(2) & 4$e$ & $3m.$ & \multicolumn{3}{c}{0} & \multicolumn{2}{c}{0} & 0.15163(8) & \multicolumn{2}{c}{1} &
\multicolumn{2}{c}{23(9)} & \multicolumn{2}{c}{$U_{11}$} & \multicolumn{2}{c}{50(16)} & 
\multicolumn{3}{c}{$\frac{1}{2}U_{11}$} & \multicolumn{2}{c}{0} & \multicolumn{2}{c}{0} & 
\multicolumn{2}{c}{47(3)} \\

O(3) & 12$k$ & $.m.$ & \multicolumn{3}{c}{0.1526(1)} & \multicolumn{2}{c}{0.3052(2)} & 0.05381(5)  & \multicolumn{2}{c}{1} & 
\multicolumn{2}{c}{56(3)} & \multicolumn{2}{c}{$U_{11}$} & \multicolumn{2}{c}{54(4)} & 
\multicolumn{3}{c}{17(4)} & \multicolumn{2}{c}{9(2)} & \multicolumn{2}{c}{$-U_{13}$} & 
\multicolumn{2}{c}{60(3)} \\

O(4) & 12$k$ & $.m.$ & \multicolumn{3}{c}{0.50585(10)} & \multicolumn{2}{c}{0.01170(10)} & 0.15208(4) & \multicolumn{2}{c}{1} & 
\multicolumn{2}{c}{36(3)} & \multicolumn{2}{c}{$U_{11}$} & \multicolumn{2}{c}{63(4)} & 
\multicolumn{3}{c}{15(3)} & \multicolumn{2}{c}{-4(1)} & \multicolumn{2}{c}{$-U_{13}$} & 
\multicolumn{2}{c}{46(3)} \\

O(5) & 4$f$ & $3m.$ & \multicolumn{3}{c}{$\frac{1}{3}$} & \multicolumn{2}{c}{$\frac{2}{3}$} & 0.55856(9) & \multicolumn{2}{c}{1} & 
\multicolumn{2}{c}{33(4)} & \multicolumn{2}{c}{$U_{11}$} & \multicolumn{2}{c}{74(8)} & 
\multicolumn{3}{c}{$\frac{1}{2}U_{11}$} & \multicolumn{2}{c}{0} & \multicolumn{2}{c}{0} & 
\multicolumn{2}{c}{47(4)} \\ \hline                               
\end{tabular}
\end{table*}

\subsection{Magnetometry}

\begin{figure*}[t]
    \centering

    \begin{minipage}{0.48\textwidth}
        \centering
        \includegraphics[width=\linewidth]{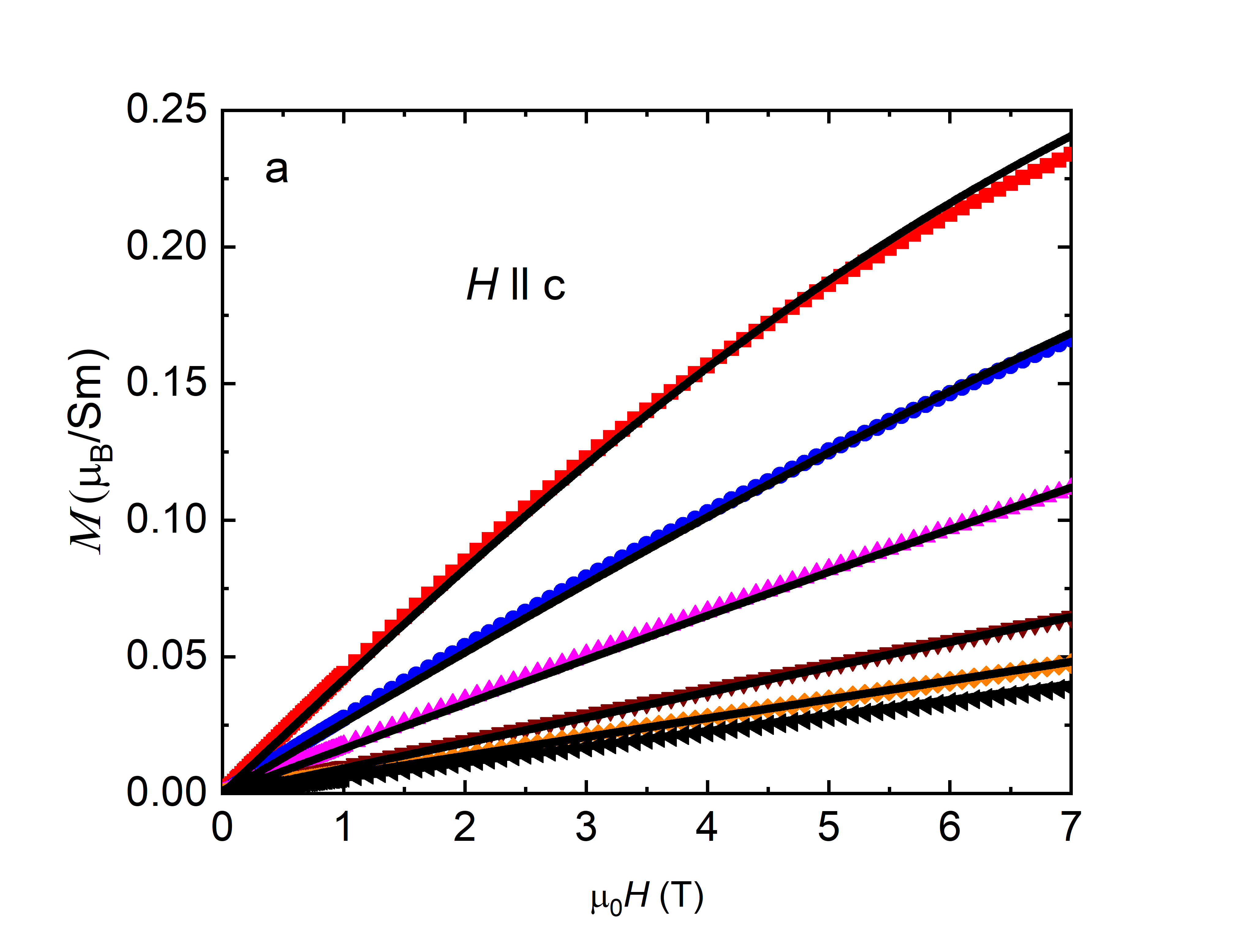}
    \end{minipage}
    \hfill
    \begin{minipage}{0.48\textwidth}
        \centering
        \includegraphics[width=\linewidth]{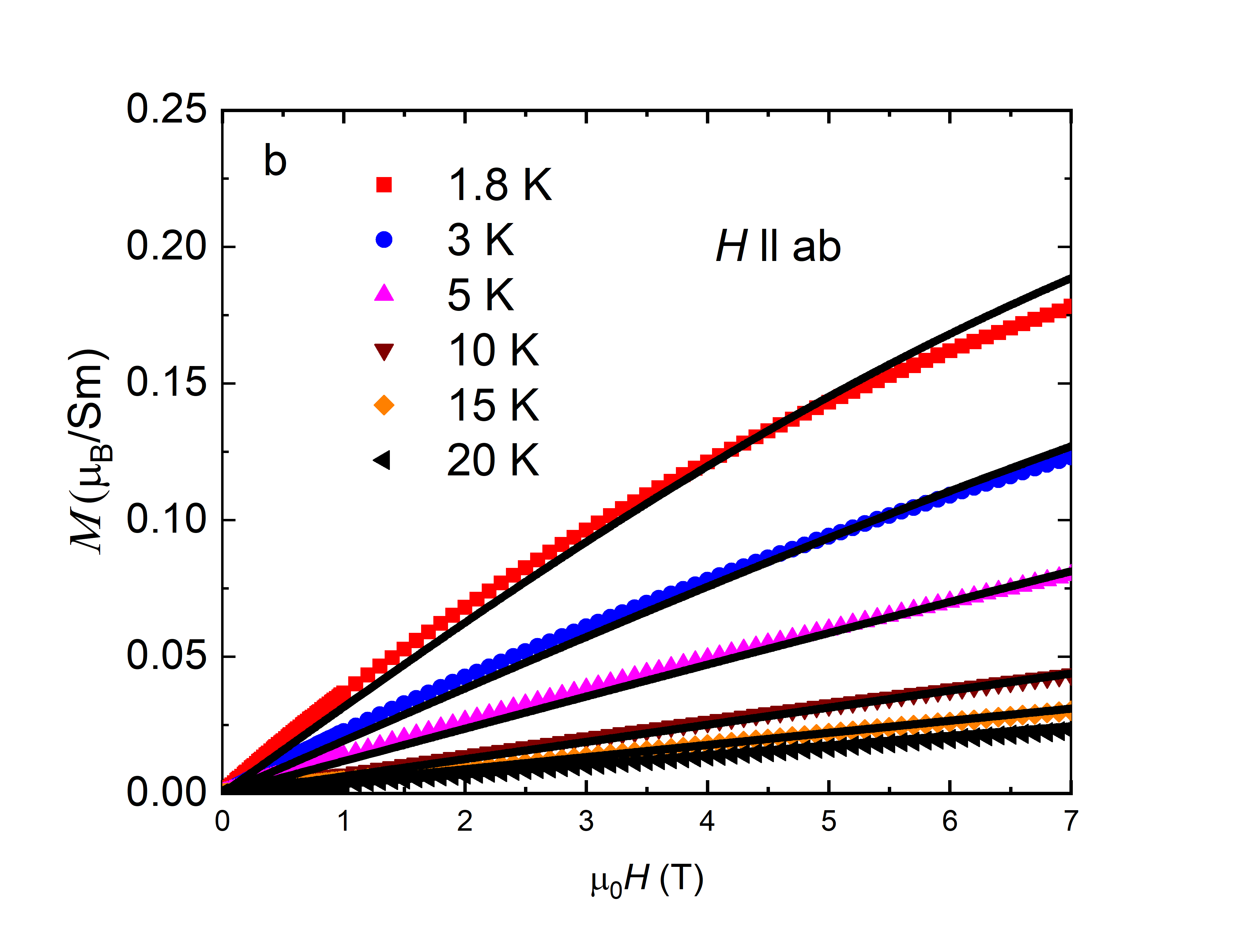}
    \end{minipage}

    \caption{
    (a) Isothermal magnetization $M(H)$ of SmMgAl$_{11}$O$_{19}$ for 
    $H \parallel c$ at several temperatures.
    (b) Isothermal magnetization $M(H)$ for $H \parallel ab$.
    Both panels show the weak anisotropy and absence of saturation up to 7~T.
    }
    \label{fig:Fig3}
\end{figure*}

Magnetic susceptibility data were first analyzed over the temperature range
1.8--100~K [Fig.~\ref{fig:Fig2}].
 In both orientations, $\chi^{-1}(T)$ exhibits strong curvature and no extended
linear Curie--Weiss (CW) regime, indicating a substantial temperature-independent
Van Vleck contribution from virtual admixture with low-lying electronic excited states, as is typical of Sm-compounds~\cite{Bairwa2025SmTa7O19,PhysRevB.99.134415}. Nevertheless, fitting the data using a Curie--Weiss expression with
temperature-independent terms,
\begin{equation}
\chi(T)=\frac{C}{T-\Theta_{\mathrm{CW}}}+\chi_{0},
\label{eq:CW}
\end{equation}

yields consistently small Weiss temperatures in both orientations, confirming an
extremely weak mean-field
exchange interactions (Fig.~\ref{fig:Fig2}b).For $H \parallel ab$, we obtain
$\theta_{\mathrm{CW}}^{ab} = 0.15(6)$~K and
$\mu_{\mathrm{eff},ab} = 0.51~\mu_{\mathrm{B}}$.
For $H \parallel c$, the corresponding fit gives
$\theta_{\mathrm{CW}}^{c} = 0.68(1)$~K and
$\mu_{\mathrm{eff},c} = 0.59~\mu_{\mathrm{B}}$.
Interpreting these as moments of an effective $S_{\mathrm{eff}} = \tfrac{1}{2}$
doublet, using
$\mu_{\mathrm{eff}} = g \sqrt{S_{\mathrm{eff}}(S_{\mathrm{eff}}+1)}\,\mu_{\mathrm{B}}
= g \tfrac{\sqrt{3}}{2}\,\mu_{\mathrm{B}}$,
yields ground-state $g$ factors
$g_{ab} \simeq 0.59$ and $g_{c} \simeq 0.68$.
These small effective moments and enhanced $g$ factors, which differ markedly from the $^6H_{\frac{5}{2}}$ ground-multiplet free-ion expectations for Sm$^{3+}$ ($g_J = 2/7$ and $\mu_{\mathrm{eff}} = g_J\sqrt{J(J+1)} \approx 0.85 \mu_B$), highlight the importance of virtual mixing with the low-lying excited $J$-multiplet in renormalising the ground-state magnetic properties~\cite{AbragamBleaney1970}.

A more reliable description of the magnetic response is obtained by restricting the fit
to the low-temperature regime, where only the ground-state Kramers doublet is expected to be
thermally populated (Fig.~\ref{fig:Fig2}c). For $H \parallel ab$, a Curie--Weiss fit
including a temperature-independent term, $\chi(T)=C/(T-\Theta_{\mathrm{CW}})+\chi_0$,
between 1.8 and 9.5~K yields
$\theta_{\mathrm{CW}}^{ab} = 0.08(8)$~K,
$\mu_{\mathrm{eff},ab} = 0.56(2)\,\mu_{\mathrm{B}}$,
and $\chi_{0}^{ab} = 2.21\times10^{-4}$~emu/mol.
For $H \parallel c$, the corresponding fit gives
$\theta_{\mathrm{CW}}^{c} = 2.8(1.9)\times10^{-4}$~K,
$\mu_{\mathrm{eff},c} = 0.60(2)\,\mu_{\mathrm{B}}$,
and a markedly larger background
$\chi_{0}^{c} = 1.56\times10^{-3}$~emu/mol.
Interpreting these effective moments in terms of an
$S_{\mathrm{eff}}=\tfrac{1}{2}$ doublet yields ground-state $g$ factors
$g_{ab}\simeq 0.65$ and $g_{c}\simeq 0.70$.

The reduced effective moments and the strong anisotropy of $\chi_{\mathrm{vv}}$ are consistent with a Kramers doublet ground state whose magnetic response is strongly renormalised by inter-multiplet ($J$-mixing) effects, leading to moderate axial anisotropy and extremely weak effective exchange interactions.

Isothermal magnetization measurements reveal a pronounced but unsaturated response
at low temperature in both field orientations (Fig.~\ref{fig:Fig3}). At 1.8~K,
$M(H)$ shows clear curvature (sublinear increase with field) and reaches
$M_{c}\approx0.24~\mu_{\mathrm{B}}/\mathrm{Sm}$ for $H\parallel c$ and
$M_{ab}\approx0.19~\mu_{\mathrm{B}}/\mathrm{Sm}$ for $H\parallel ab$ at the highest
applied field, without any sign of saturation. Upon warming to 3~K and 5~K, the
overall magnetization is reduced across the full field range. By 10~K the
magnetization becomes essentially linear in both directions and the anisotropy
between $H\parallel c$ and $H\parallel ab$ is strongly diminished.

To quantify the low-energy response in terms of an effective Kramers doublet, we
fit $M(H)$ using a modified spin-$\tfrac{1}{2}$ Brillouin form,
\begin{equation}
M(H)=M_{\mathrm{sat}}\,\tanh(x)+\chi_0\,\mu_0 H,
\label{eq:M_brillouin}
\end{equation}
\begin{equation}
x=\frac{g\,\mu_{\mathrm{B}}\mu_0 H}{2k_{\mathrm{B}}T},
\label{eq:x_def}
\end{equation}
where $g$ is the effective $g$ factor of the ground doublet and
$M_{\mathrm{sat}}=N_{\mathrm{A}}g\mu_{\mathrm{B}}/2$ for one effective spin-$\tfrac{1}{2}$
per Sm. In both orientations, agreement with Eqs.~(\ref{eq:M_brillouin}) and
(\ref{eq:x_def}) becomes apparent only for $T\ge 5$~K, while clear systematic
deviations persist at 1.8~K and 3~K (more pronounced for $H\parallel ab$).
Fitting the data for $H\parallel c$ yields $g_{c}=0.65213(36)$ and
$\chi_{0,c}=2.14(2)\times 10^{-3}$, whereas for $H\parallel ab$ we obtain
$g_{ab}=0.57628(69)$ and $\chi_{0,ab}=7.31(33)\times 10^{-4}$.

The breakdown at the lowest temperatures is consistent with additional low-energy
contributions beyond an isolated two-level description.
The temperature-independent background extracted from $M(H)$ fits is of the same
order as that obtained from Curie--Weiss analysis of $\chi(T)$, with modest
differences likely reflecting the distinct sensitivity of the two protocols to
field-linear contributions and weak non-idealities (e.g., internal-field
distributions and fitting-window dependence).

\subsection{Crystal electric field calculations}
\label{sec:cef}

To quantify the single-ion anisotropy of Sm$^{3+}$ in SmMgAl$_{11}$O$_{19}$, we
performed CEF calculations using the
\textsc{PyCrystalField} package~\cite{Scheie2021} and the refined
crystallographic structure. Sm$^{3+}$ occupies a unique rare-earth site in the
magnetoplumbite lattice and is coordinated by twelve oxygen ligands forming a
distorted SmO$_{12}$ polyhedron with approximate trigonal symmetry. The local
axes were chosen as $z\parallel c$ and $y\parallel a$, and the ligand
environment was truncated at $r\le 3$~\AA, which captures the twelve nearest
oxygen ions. To avoid ambiguity from crystallographic multiplicity, we
explicitly selected the Sm(1) site (occupancy 1) when constructing the local
environment. The twelve oxygen ligands form two closely spaced radial shells at
$d\simeq 2.641$~\AA\ and $2.793$~\AA.

As a baseline electrostatic estimate we adopted a purely ionic point-charge
model in which the oxygen ligands were assigned $q_{\mathrm O}=-2e$ while all
non-oxygen neighbors were taken as neutral. The resulting single-ion Hamiltonian
can be expressed in Stevens-operator form,
\begin{equation}
  \mathcal{H}_{\mathrm{CEF}}=\sum_{n,m} B_m^{n} O_m^{n},
\end{equation}
where $O_m^{n}$ are Stevens operators and $B_m^{n}$ are the CEF parameters. With
the above axis choice and the near-trigonal ligand geometry, the point-charge
solution yields the dominant trigonal terms ($B_2^0$, $B_4^0$, $B_6^0$, and
$B_6^6$), while the remaining coefficients are negligible within numerical
precision.

A CEF description restricted to a single $J$ multiplet can be incomplete for
Sm$^{3+}$ because the temperature-independent susceptibility is often dominated
by virtual excitations and intermultiplet (``$J$-mixing'') processes. We
therefore diagonalized the single-ion problem in the intermediate-coupling
(LS-mixed) basis implemented in \textsc{PyCrystalField}. In this framework, the
parameter \texttt{LS\_Coupling} controls the degree of multiplet mixing and
provides a phenomenological way to include intermultiplet contributions, which
are well documented in Sm$^{3+}$ magnetism (see, e.g.,
Ref.~\cite{PecanhaAntonio2019_SmPyrochlores}). In our analysis, we used the
experimentally determined Van~Vleck susceptibility
as an empirical constraint to optimize the multiplet-mixing strength in the
point-charge model and thereby renormalize the low-energy response into
quantitative agreement with experiment. The results quoted below correspond to
\texttt{LS\_Coupling}$=17$~meV (for the justification for this value, see Discussion).

Within this LS-mixed ionic point-charge model, the low-energy CEF spectrum
consists of a Kramers doublet ground state followed by excited doublets at
$\Delta_1 = 7.34$~meV and $\Delta_2 = 25.01$~meV (relative to the ground
doublet), with the next levels above $\sim 60$~meV. These higher-lying states
are not thermally populated at low temperature but contribute to the response
through virtual excitations (Van~Vleck channel).

The ground-doublet $g$ tensor in the chosen local frame is nearly diagonal but
contains a small $xz$ mixing,

\[
\mathbf{g}=\begin{pmatrix}
0.509 & 0 & 0.330\\
0 & 0.607 & 0\\
0.381 & 0 & 0.587
\end{pmatrix}.
\]

corresponding to principal values
$(g_1,g_2,g_3)\approx(0.607,\,0.607,\,0.700)$. Identifying the principal axis
closest to the crystallographic $c$ direction as $g_c$ yields
$g_{ab}\simeq 0.6$ and $g_c\simeq 0.7$, in excellent agreement with the
$g\approx0.6$--$0.7$ values extracted from low-temperature magnetometry.

We emphasize that the calculation is still based on an ionic point-charge
description and therefore does not explicitly include covalency or screening,
so a spectroscopic determination of the CEF levels is ultimately required for a
confident assignment.

Nevertheless, for Sm$^{3+}$ the dominant correction to a single-$J$ description
arises from intermultiplet admixture, which is naturally captured by the
LS-mixed framework. By calibrating the degree of multiplet mixing using the
experimental Van~Vleck contribution, our optimized
point-charge model achieves quantitative agreement with the experimentally
inferred $g$ tensor and supports an effective pseudospin-$\tfrac{1}{2}$ Kramers
doublet below $\sim10$~K, with the next CEF excitations on energy scales far
above the exchange scale inferred from thermodynamic measurements.

\subsection{Specific heat and magnetic entropy}

\begin{figure*}
    \centering

 \includegraphics[width=0.48\textwidth]{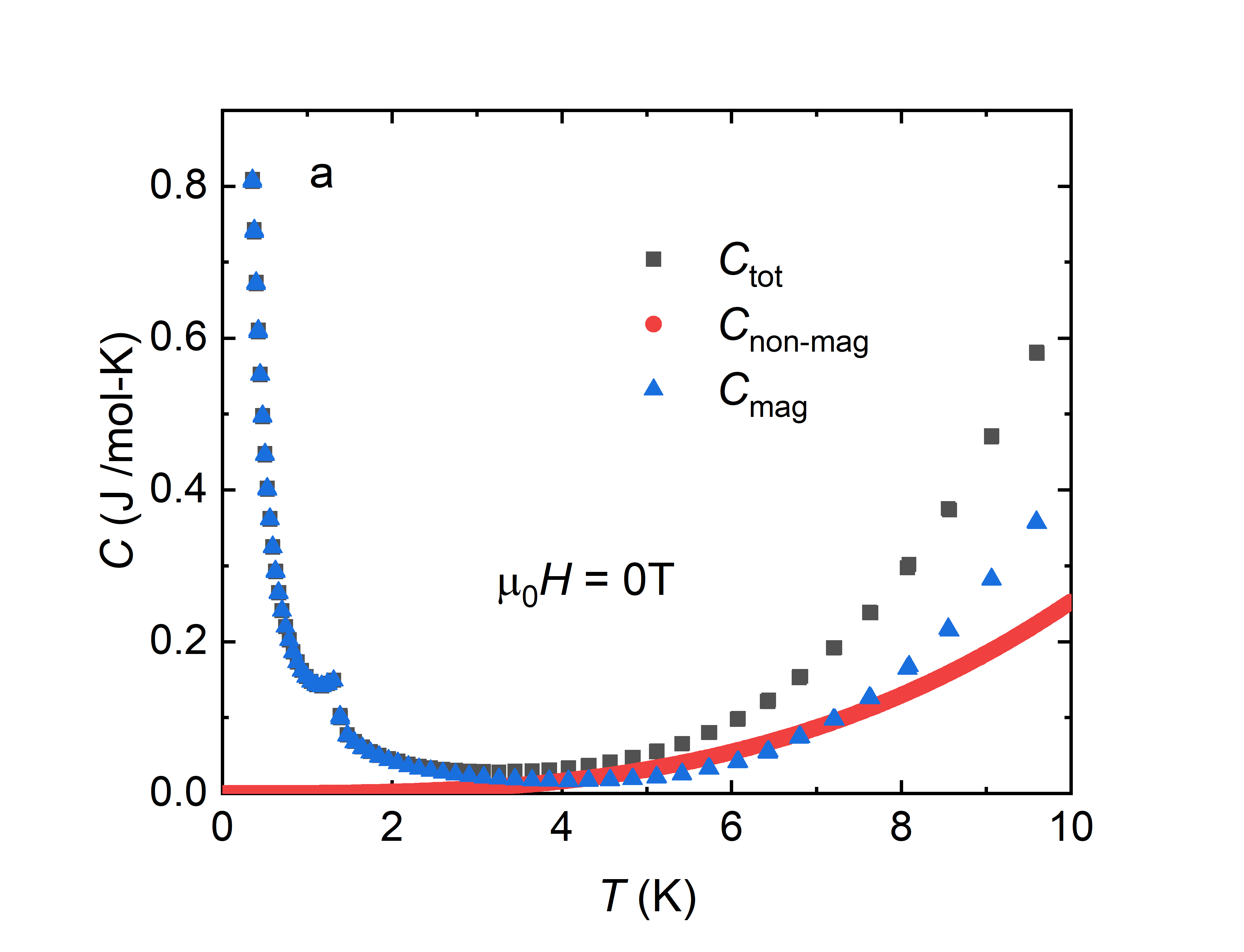}\hfill
    \includegraphics[width=0.48\textwidth]{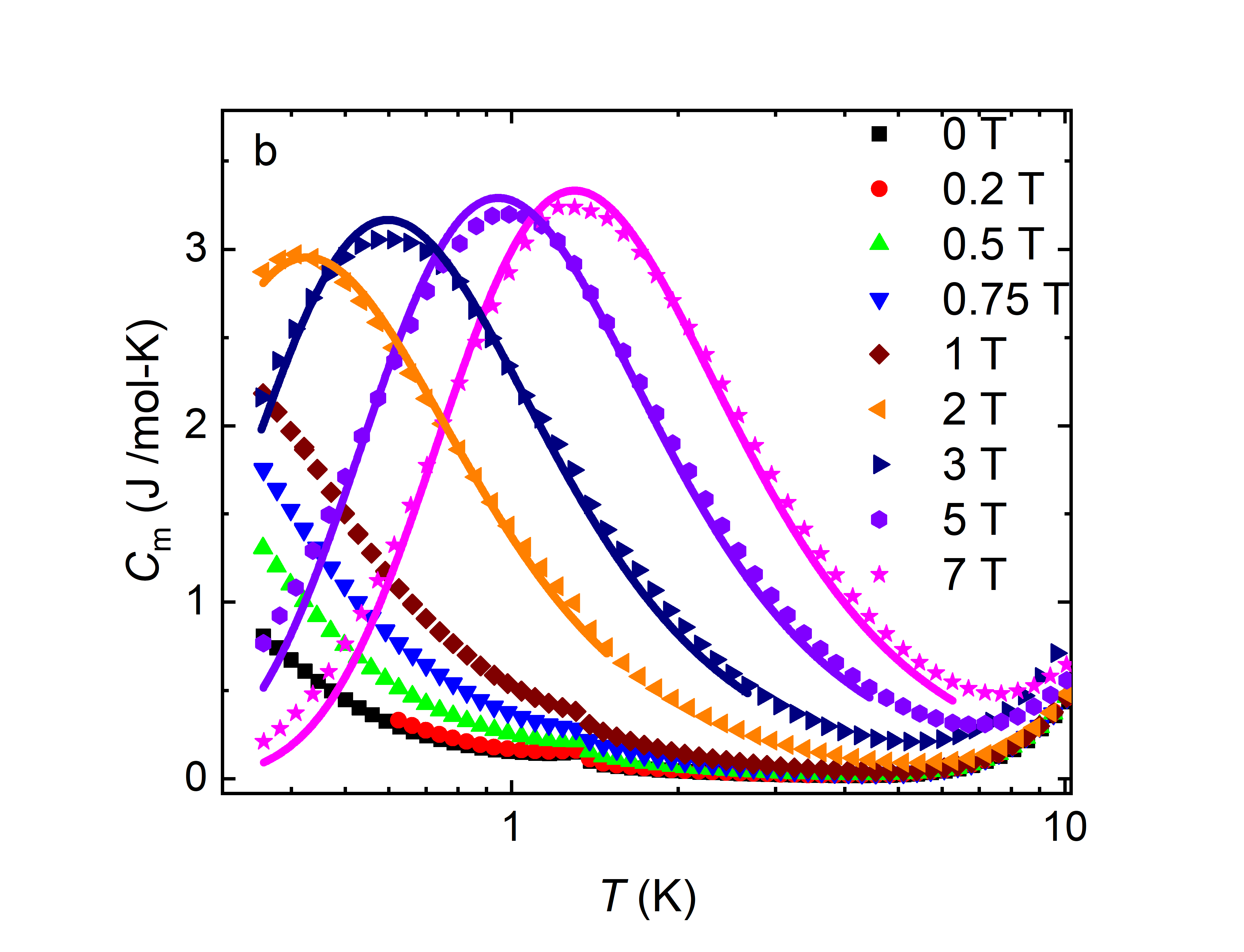}\\[1.5mm]

    \includegraphics[width=0.48\textwidth]{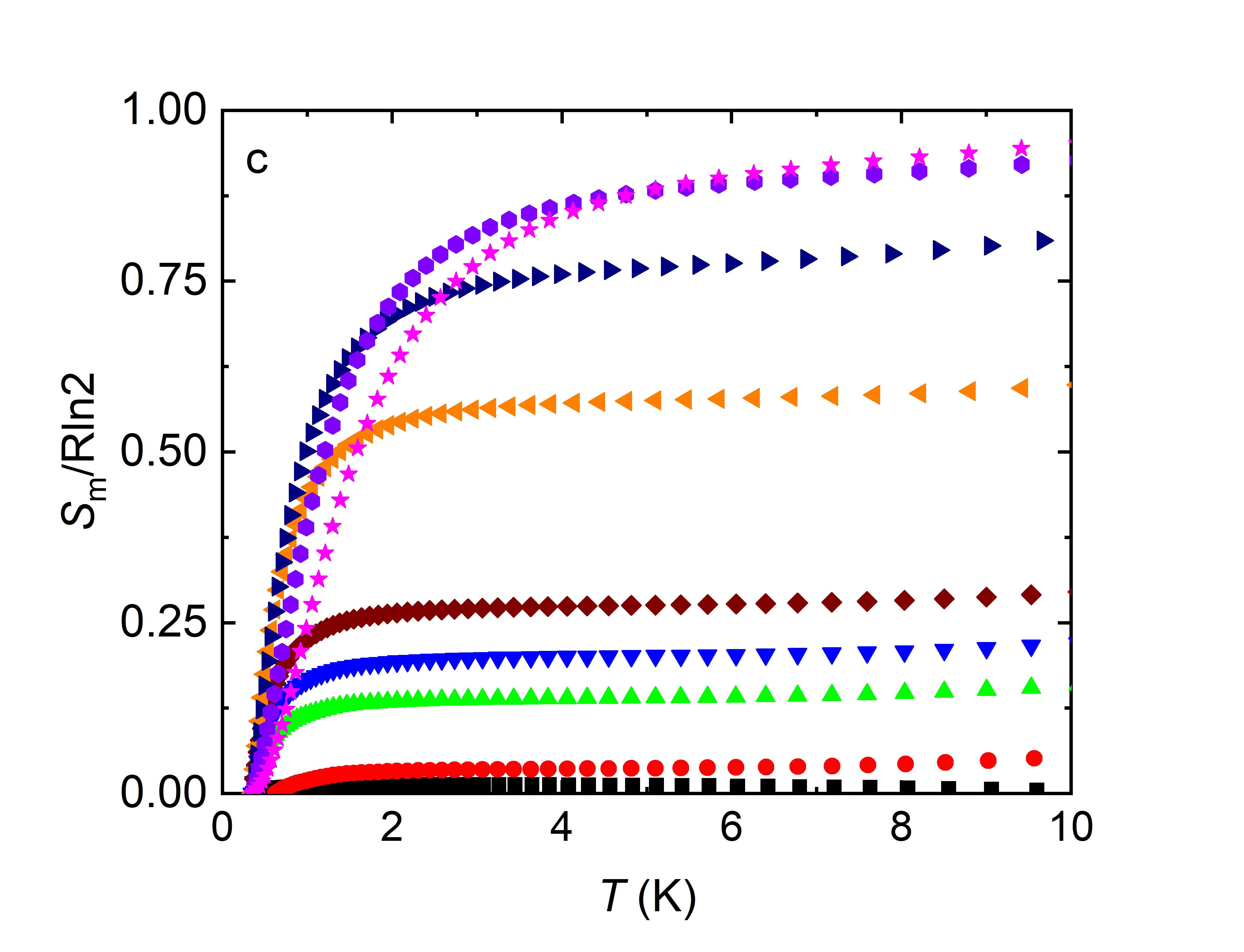}\hfill
    \includegraphics[width=0.48\textwidth]{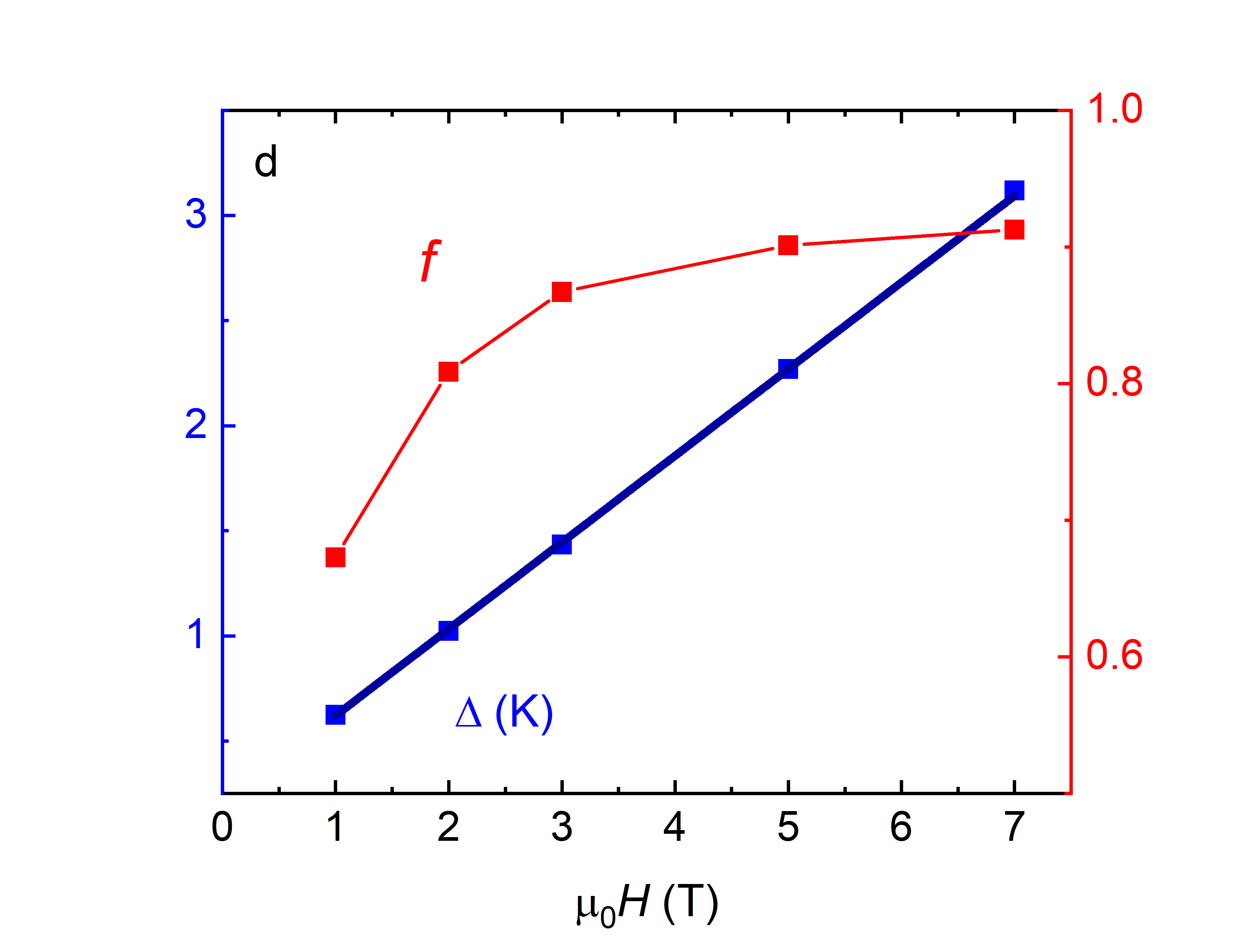}

    \caption{
    \textbf{Fig.~3.} Specific heat and Schottky analysis of SmMgAl$_{11}$O$_{19}$.
    (a) Total specific heat $C_p(T)$ of SmMgAl$_{11}$O$_{19}$ together with the
    nonmagnetic analogue LaMgAl$_{11}$O$_{19}$, and the magnetic contribution
    $C_m(T)$ extracted by subtracting the phonon background in zero field.
    (b) Magnetic specific heat $C_m(T)$ of SmMgAl$_{11}$O$_{19}$ under various applied
    magnetic fields. Solid curves show two-level Schottky fits for $\mu_0H \ge 1$~T.
    (c) Magnetic entropy $S_m(T)$ obtained from integrating $C_m/T$.
    (d) Zeeman gap $\Delta(B)$ and filling fraction $f(B)$ extracted from the Schottky
    analysis. The solid blue line represents a linear fit to $\Delta(B)$.
    }

    \label{fig:Fig4}
\end{figure*}

The specific-heat data provide a complementary thermodynamic perspective on the
low-energy physics of SmMgAl$_{11}$O$_{19}$. In zero field, the magnetic specific heat
$C_m(T)$ shows no $\lambda$-type anomaly down to 0.35~K, excluding conventional
long-range magnetic order within our experimental window (Fig.~\ref{fig:Fig4}a). Instead,
$C_m(T)$ exhibits a low-temperature upturn with a weak kink-like feature. Given the
absence of any corresponding anomaly in susceptibility or magnetization, these
low-temperature features are most naturally attributed to a combination of nuclear
Schottky contributions from Sm isotopes and minor impurity- or defect-related effects.

Applying a magnetic field drives the system toward a regime dominated by single-ion
physics. For fields of 2--7~T, $C_m(T)$ develops a pronounced Schottky-like peak that
shifts systematically to higher temperature with increasing field, as expected for a
Zeeman-split Kramers doublet (Fig.~\ref{fig:Fig4}a). These anomalies are well described
by a simple two-level Schottky form
\begin{equation}
C_{\mathrm{Sch}}
 = f\,R \left(\frac{\Delta}{T}\right)^{2}
   \frac{\exp\!\big[-\Delta/T\big]}
        {\big[1+\exp\!\big(-\Delta/T\big)\big]^{2}},
\label{eq:Schottky}
\end{equation}
where $R$ is the molar gas constant, $\Delta$ is the field-dependent energy gap of the
effective Kramers doublet, and $f$ is an effective filling fraction that accounts for the
reduced number of active Sm moments.
 The gaps extracted from these
high-field anomalies increase linearly with field and correspond to an
out-of-plane effective $g_c \simeq 0.62$, in excellent
agreement with the magnetization-based estimates. 
At lower fields, the evolution of the Schottky anomaly reveals the emergence of a regime
where magnetic correlations play a more significant role in the low temperature properties. When
the field is reduced below $\sim 3$~T, the peak amplitude of the hump is markedly
suppressed, in contrast to the field-independent peak height expected for an ideal
two-level non interacting Schottky anomaly with fixed entropy (Fig.~\ref{fig:Fig4}a). In the language of
the Schottky fits, this corresponds to a progressive reduction of the effective filling
fraction $f$ with decreasing field, with $f$ dropping particularly strongly already at
$\mu_{0}H = 1$~T (Fig.~\ref{fig:Fig4}c). Such amplitude loss implies that only a subset of
Sm$^{3+}$ ions experiences a well-defined Zeeman gap at low fields, while the remaining
moments contribute to a broad distribution of low-energy configurations shaped by weak
exchange, dipolar interactions, and disorder on the frustrated triangular lattice.
 This interpretation is consistent with the sizable residual entropy that persists down to
the base temperature at low and zero field, and with the anisotropic magnetic response
inferred from susceptibility and magnetization. The field dependence of the entropy further
underscores this picture: only in sufficiently strong fields does the system fully unlock
the doublet degrees of freedom, with the magnetic entropy approaching the full $R\ln 2$
between 0.35 and 10~K by $\mu_{0}H \approx 5$~T (Fig.~\ref{fig:Fig3}b). At the same time, Despite the Curie--Weiss temperatures remaining very close to zero, the Schottky fits
reveal a strong suppression of the effective filling fraction $f$ as the field is
reduced, indicating that a substantial fraction of Sm$^{3+}$ moments is effectively
quenched rather than simply exchange-decoupled. This is fully compatible with the
reduced effective moment extracted from magnetization and with the sizable
Van~Vleck contributions to the susceptibility in both field directions. Taken together,
the thermodynamic data establish a crossover from a low-field correlated regime with
entropy locked in low-energy modes to a high-field single-ion regime governed by a
Zeeman-split Kramers doublet.

\begin{figure*}[t]
    \centering

    \begin{minipage}{0.48\textwidth}
        \centering
        \includegraphics[width=\linewidth]{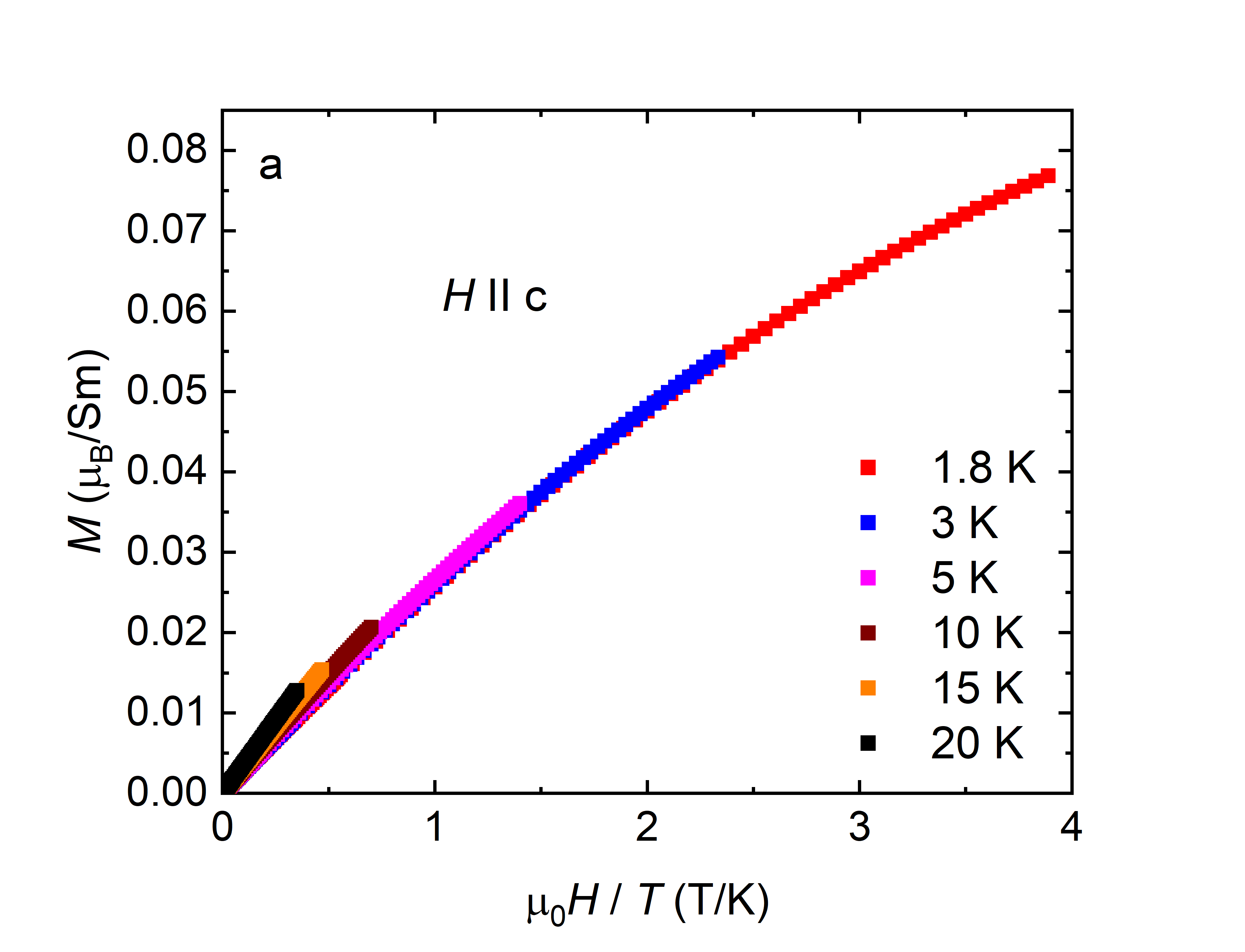}
    \end{minipage}
    \hfill
    \begin{minipage}{0.48\textwidth}
        \centering
        \includegraphics[width=\linewidth]{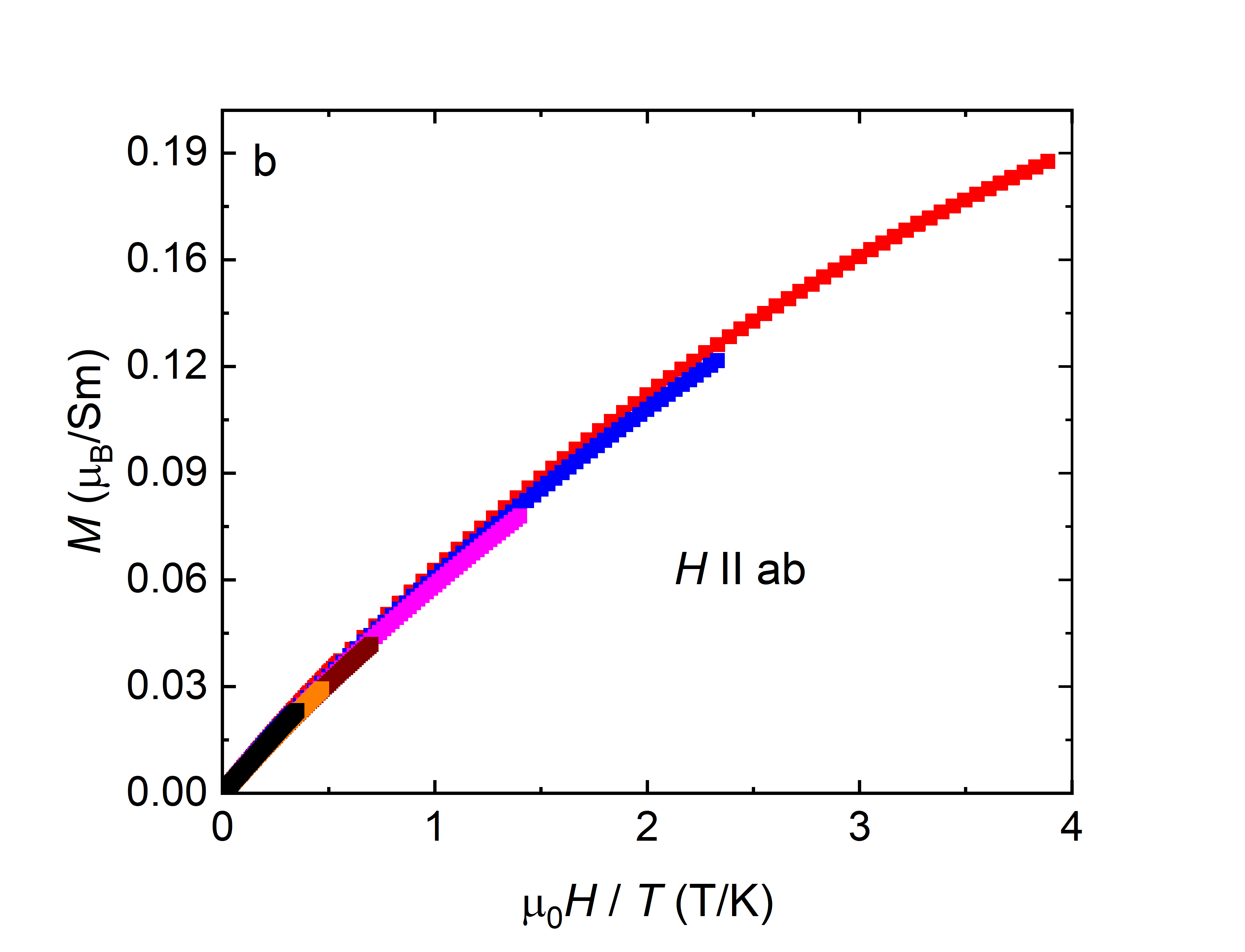}
    \end{minipage}

    \caption{
    (a) Scaling plot of magnetization for SmMgAl$_{11}$O$_{19}$ with 
    magnetization $M$ plotted against $H/T$ for fields applied along 
    the $c$ axis. The absence of scaling collapse indicates the presence 
    of additional low-energy correlations and internal fields for 
    $H \parallel c$.
    (b) Scaling of $M$ vs.\ $H/T$ for $H \parallel ab$. 
    The near-universal collapse of the curves is consistent with 
    predominantly single-ion paramagnetic behavior in the $ab$ plane.
    }
    \label{fig:Fig5}
\end{figure*}

\section{Discussion}

The magnetometry of SmMgAl$_{11}$O$_{19}$ places the material in the weak-exchange
limit of a rare-earth triangular-lattice magnet in which single-ion physics dominates,
yet the consequences of weak residual interactions remain experimentally visible. Curie--Weiss analysis
yields Weiss temperatures that are essentially zero in both orientations, demonstrating
that the net exchange scale is extremely small. Using a mean-field estimate for a
triangular lattice ($z = 6$) with an effective $S = \tfrac{1}{2}$ doublet, the magnitude
of a nearest-neighbor exchange compatible with such tiny $\Theta_{\mathrm{CW}}$ is on the
order of $|J|/k_{\mathrm{B}} \lesssim 0.05$~K, i.e., far below typical crystal-field and
Zeeman energy scales. In this regime, geometric frustration, together with the
$\sim 6\%$ Sm-site deficiency, can efficiently cancel and randomize weak interactions,
leaving the response governed primarily by a strongly quenched local Kramers doublet.
At the same time, the absence of a sizable exchange scale means that even subtle
disorder- or interaction-induced renormalizations remain visible in the low-temperature
response.

\begin{figure*}[t]
    \centering

    \begin{minipage}{0.48\textwidth}
        \centering
        \includegraphics[width=\linewidth]{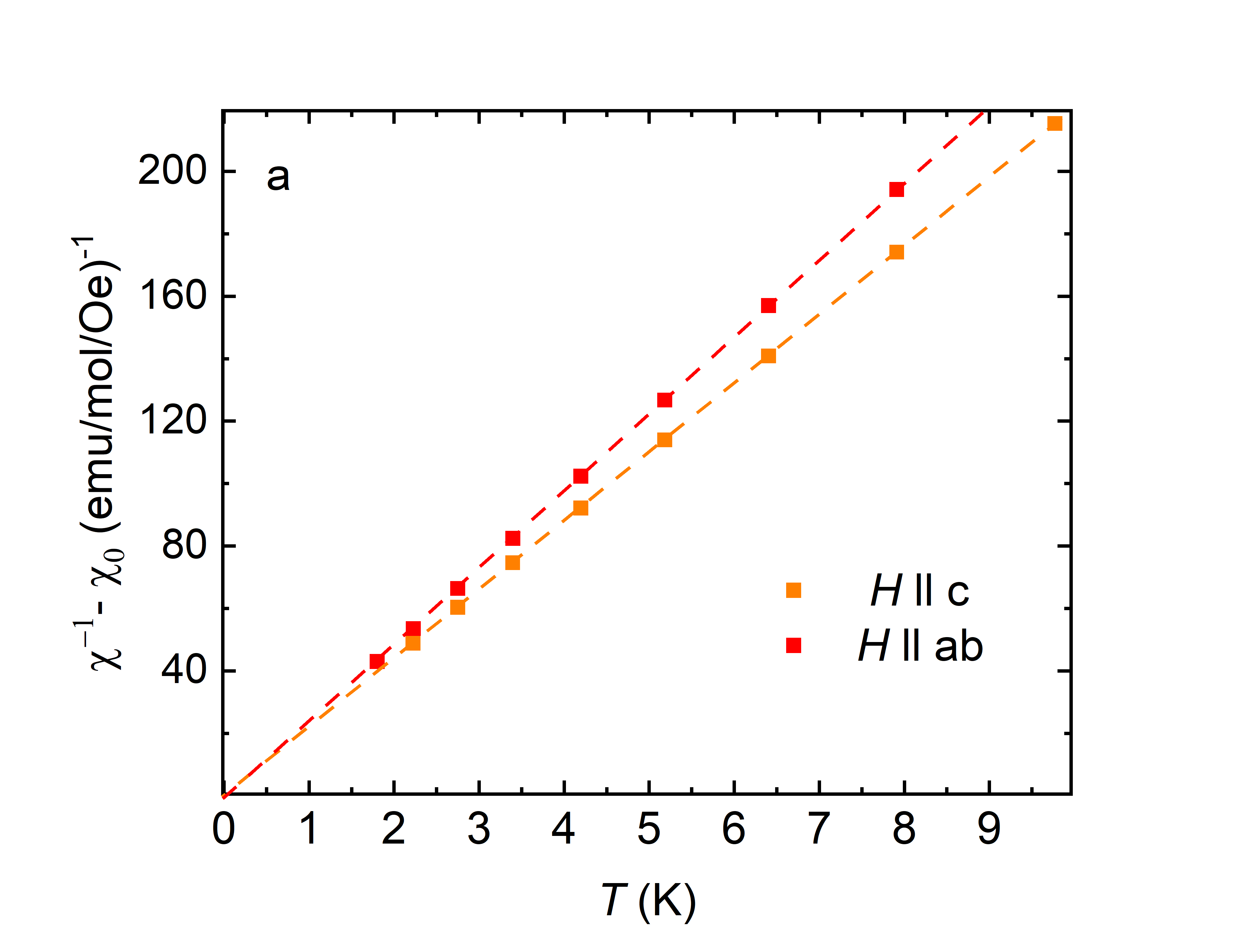}
    \end{minipage}
    \hfill
    \begin{minipage}{0.48\textwidth}
        \centering
        \includegraphics[width=\linewidth]{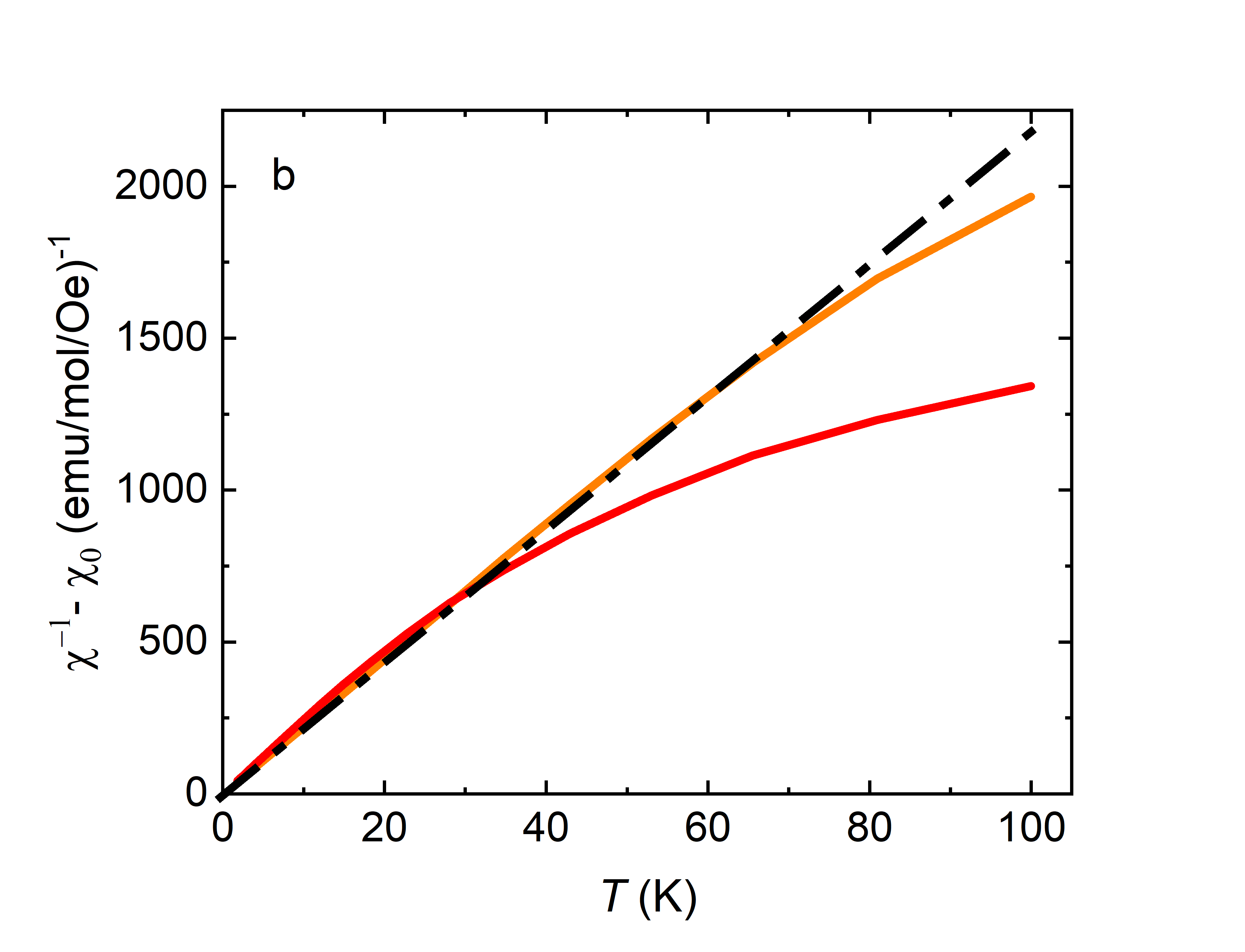}
    \end{minipage}

    \caption{
    (a) Inverse corrected susceptibility $(\chi - \chi_{0})^{-1}$ as a function of
    temperature in the low-temperature window $1.8 \leq T \leq 10$~K, showing
    a nearly linear Curie--Weiss behaviour of the ground-state Kramers doublet.
    (b) $(\chi - \chi_{0})^{-1}$ over the extended temperature range, highlighting
    the breakdown of simple Curie--Weiss scaling.
    }
    \label{fig:Fig6}
\end{figure*}

Within this weak-exchange background, the low-temperature susceptibility reveals a clear
directional contrast. In the $ab$ plane, $M/H$ is well described by a Curie law once the
small, temperature-independent Van~Vleck contribution is taken into account, and the
slight sub-Curie trend in the raw $M/H$ mainly reflects this additive
$\chi_{\mathrm{vv}}^{ab}$. Consistent with an almost single-ion Kramers doublet, the
isothermal magnetization for $H \parallel ab$ collapses onto an approximately universal
curve when plotted as $M$ against $H/T$ (Fig.~\ref{fig:Fig5}b), as expected for a
paramagnet with a single dominant energy scale set by the Zeeman splitting. For both
field directions, the corrected inverse susceptibility $(\chi - \chi_0)^{-1}$ is nearly
linear between 1.8 and 10~K, showing that in this low-temperature window the response is
well captured by a Curie term from an effective doublet plus a temperature-independent
background (Fig.~\ref{fig:Fig6}a). Interpreting the effective moments in terms of an
$S_{\mathrm{eff}} = \tfrac{1}{2}$ doublet yields ground-state $g$ factors
$g_{ab} \simeq 0.65$ and $g_{c} \simeq 0.70$, reduced compared to the
free-ion Sm$^{3+}$ value and already indicating strong quenching of the local moment.

When the same analysis is extended to higher temperatures, clear deviations from
this simple picture emerge. Extending $(\chi - \chi_0)^{-1}$ up to 100~K, we find that
for $H \parallel ab$ the corrected inverse susceptibility bends away from linearity
already above $T \sim 20$~K, while for $H \parallel c$ a noticeable deviation sets in
above $T \sim 80$~K (Fig.~\ref{fig:Fig6}b). This demonstrates that the quenching of the Sm$^{3+}$ moment cannot
be attributed to a single mechanism such as a rigid, isolated CEF doublet with a fixed
$g$ tensor. Instead, it reflects a combination of crystal-field splitting within the
$J = 5/2$ manifold, admixture of the excited $J = 7/2$ multiplet, and weak residual
interactions and disorder, whose relative importance grows as thermally populated
excited states contribute to the response.

To quantify the role of $J$-multiplet mixing, we interpret the Van~Vleck term in the
standard $J$-mixing framework for Sm$^{3+}$~\cite{Wagner1972_TheoryOfMagnetism}. In this approach the
temperature-independent susceptibility is written as
\begin{equation}
  \chi_{\mathrm{vv}} = \frac{N_A \mu_B^2}{k_B}\,\alpha_J,
\end{equation}
with
\begin{equation}
  \alpha_J = \frac{20}{7\Delta},
\end{equation}
where $\Delta = E_{J+1} - E_J$ is the energy separation (in kelvin) between the excited
($J = \tfrac{7}{2}$) and ground ($J = \tfrac{5}{2}$) multiplets of Sm$^{3+}$. 
Using the cgs conversion
\[
\frac{N_A \mu_B^2}{k_B} \simeq 0.375~\mathrm{emu\,K/mol},
\]

and our low-temperature fit for $H \parallel c$,
$\chi_{\mathrm{vv}}^{c} = 1.56\times10^{-3}~\mathrm{emu/mol}$, we obtain
\[
  \alpha_J^{(c)}
  = \frac{\chi_{\mathrm{vv}}^{c}}{N_A \mu_B^2/k_B}
  = \frac{1.56\times10^{-3}}{0.375}
  \simeq 4.2\times10^{-3}~\mathrm{K^{-1}},
\]
and hence an effective multiplet gap
\[
  \Delta_{\mathrm{eff}}
  = \frac{20}{7\,\alpha_J^{(c)}}
  \simeq 6.8\times10^{2}~\mathrm{K}.
\]
Expressed in energy units, this corresponds to

\[  \Delta_{\mathrm{eff}} \simeq
  \frac{6.8\times10^{2}~\mathrm{K}}{11.6045~\mathrm{K/meV}}
  \simeq 59~\mathrm{meV}.\]

To connect this estimate to the intermediate-coupling calculations, we relate the
$J=\tfrac{5}{2}\!\rightarrow\!\tfrac{7}{2}$ separation to the effective spin--orbit
(LS) coupling constant $\zeta$ via the Russell--Saunders expression
\begin{equation}
  E_J=\frac{\zeta}{2}\left[J(J+1)-L(L+1)-S(S+1)\right],
\end{equation}
with $L=5$ and $S=\tfrac{5}{2}$ for the ${}^{6}H_J$ term of Sm$^{3+}$. This gives
\begin{equation}
  \Delta_{7/2-5/2}=E_{7/2}-E_{5/2}=\frac{7}{2}\,\zeta,
\end{equation}
and therefore
\[
  \zeta_{\mathrm{eff}}=\frac{2}{7}\,\Delta_{\mathrm{eff}}
  \simeq \frac{2}{7}\times 59~\mathrm{meV}
  \simeq 16.9~\mathrm{meV}.
\]
This value provides a quantitative basis for the multiplet-mixing parameter used in
\textsc{PyCrystalField}: we set \texttt{LS\_Coupling}$\approx 17$~meV in the CEF
calculations so that the degree of intermultiplet admixture is consistent with the
Van~Vleck response extracted from $\chi(T)$.

This effective $J=\tfrac{5}{2}\!\rightarrow\!\tfrac{7}{2}$ separation is somewhat
smaller than the free-ion spin--orbit splitting expected for Sm$^{3+}$, as anticipated
since part of the observed Van~Vleck response also originates from CEF mixing within the
$J=\tfrac{5}{2}$ manifold and from higher $J$ multiplets. We therefore interpret
$\Delta_{\mathrm{eff}}$ as a lower bound on the true multiplet gap and as a quantitative
measure that $J$-multiplet mixing makes a substantial contribution to the quenching of
the Sm$^{3+}$ moment in SmMgAl$_{11}$O$_{19}$.

The isothermal magnetization provides complementary insight. Even after including a
linear $\chi_{\mathrm{0}}H$ term, an effective spin-$\tfrac{1}{2}$ Brillouin form does
not reproduce $M(H)$ quantitatively at the lowest temperatures. This mismatch is
physically informative rather than a fitting artifact: a Brillouin+Van~Vleck model
assumes a uniform two-level system with a single $g$ tensor and no distribution of local
environments. In SmMgAl$_{11}$O$_{19}$, several effects naturally violate this
idealization. Weak frustrated interactions can still renormalize the low-$T$ field
response even when $\Theta_{\mathrm{CW}} \approx 0$; Sm-site deficiency and local
disorder broaden the CEF landscape and thus generate a spread of effective $g$ factors
and internal fields; and the proximity of excited CEF levels enables additional
field-induced mixing beyond a strict two-level description. The improvement of
Brillouin-like behavior at intermediate temperatures is then consistent with thermal
suppression of the weakest disorder- and interaction-induced effects, whereas deviations
re-emerge upon further heating as excited CEF states become thermally populated.

The thermodynamic data provide a crucial and fully consistent perspective on these
magnetic trends. Specific-heat measurements were performed with $H \parallel c$ only,
because the crystals cleave very strongly in the $ab$ plane, resulting in thin platelets
that are difficult to mount with the field perpendicular to the $c$ axis. Along the
$c$ axis, the zero- and low-field regime shows no ordering anomaly down to the base
temperature and releases only a small fraction of the doublet entropy. At low
fields, a Schottky-like anomaly develops, but with a noticeably reduced peak amplitude
compared to the high-field limit, indicating that only a subset of Sm$^{3+}$ ions
contributes to a well-defined Zeeman-split two-level response, while the remaining
spectral weight is tied up in low-energy configurations that do not behave as an ideal
Schottky doublet. In other words, the system is not fully polarized in fields below about
5~T, and a finite fraction of the doublet entropy remains locked in weakly interacting or
disordered degrees of freedom. This picture is consistent with the low-temperature
deviations of $M(H)$ from a simple Brillouin form and explains why a single effective gap
cannot describe the magnetization in the weak-field regime.

Upon increasing the field along $c$, the system crosses over into a high-field single-ion
regime dominated by a Zeeman-split Kramers doublet. The Schottky anomaly becomes well
defined, its gap grows linearly with field, and the integrated entropy approaches the
doublet value of $0.94\,R\ln 2$, demonstrating that strong fields suppress
internal-field distributions and residual correlations, thereby restoring an almost ideal
two-level response. The nearly vanishing Curie--Weiss temperatures and the reduced
effective $g$ factor ($g_c \simeq 0.6$) show that this response is governed primarily by
a strongly quenched Sm$^{3+}$ doublet and its Zeeman splitting, rather than by a sizable
exchange gap. The concurrent improvement of magnetization scaling and the approach toward
Brillouin-like behaviour in the same field range further confirm the crossover from a
correlated low-field regime with partially locked entropy to a polarized single-ion
regime dominated by Zeeman physics of a renormalized doublet.

At higher temperatures, irrespective of field direction, the response crosses over into a
CEF-mixed paramagnetic regime where excited crystal-field levels contribute strongly.
This is reflected in the curvature of $\chi^{-1}(T)$ outside the low-temperature
Curie--Weiss window and in the re-emergence of non-Brillouin behaviour in $M(H)$ above a
few kelvin. Thus, within the measured range the overall behaviour is governed not by
sharp phase boundaries but by two crossovers: (i) a field-driven crossover along $c$ from
a low-field correlated doublet regime with reduced filling fraction $f$ to a high-field
Zeeman single-ion regime where most of the doublet entropy is released, and (ii) a thermal
crossover out of the ground-doublet manifold into a CEF-mixed paramagnet. While the
specific-heat data constrain the $T$--$B$ diagram quantitatively only for $H \parallel c$,
the magnetometry indicates that the $ab$-plane channel remains closer to the single-ion
limit over a broader temperature range, consistent with its smaller Van~Vleck contribution
and more robust scaling collapse.

It is instructive to contrast SmMgAl$_{11}$O$_{19}$ with the recently reported
SmTa$_7$O$_{19}$, as both materials host Sm$^{3+}$ Kramers ions on triangular
lattices with strong spin--orbit coupling and CEF-split effective
$J_{\rm eff}=1/2$ ground states. In SmTa$_7$O$_{19}$, sizeable antiferromagnetic
exchange [$\Theta_{\rm CW}\sim-0.4$ to $-0.7$~K in both orientations] together
with strong geometric frustration places the system in a regime where collective
dynamics dominate: zero- and longitudinal-field $\mu$SR reveal persistent
low-temperature spin fluctuations, the specific heat exhibits a $\log T$
contribution together with a characteristic $C_m\propto T^2$ power law at high
fields, and the overall thermodynamic response is consistent with a gapless,
QSL-like state~\cite{Bairwa2025SmTa7O19}.

By contrast, SmMgAl$_{11}$O$_{19}$ realizes the extreme weak-exchange limit of a
Sm-based triangular lattice. Here $\Theta_{\rm CW}^{ab}$ is close to zero, the
effective moment of the ground doublet is strongly reduced, and the high-field
magnetization and Schottky anomaly are well captured by a strongly quenched
single-ion Kramers doublet with a small effective $g$ factor, renormalized by
the combined effects of crystal-field splitting and $J$-multiplet mixing. At low
fields, the Schottky peak becomes broadened and suppressed, and the doublet
entropy is only partially released within the experimental window, indicating a
weak, anisotropic low-field regime that deviates from an ideal two-level
Schottky response and likely reflects residual internal-field distributions and
non-idealities, rather than a robust exchange-driven QSL state of the type
realized in SmTa$_7$O$_{19}$.

For simplicity, all analyses were performed assuming one Sm per formula unit,
although the refined composition is $\mathrm{Sm}_{0.941}\mathrm{MgAl}_{11}\mathrm{O}_{19}$;
therefore, the intrinsic magnetic response per Sm ion may be slightly larger.

Within the broader LnMgAl$_{11}$O$_{19}$ hexaaluminate family, SmMgAl$_{11}$O$_{19}$
thus occupies the most weakly interacting, single-ion--dominated limit of the
phase diagram. CeMgAl$_{11}$O$_{19}$ hosts an XXZ-anisotropic Kramers doublet
tuned close to a quantum critical point~\cite{Bastien2025}, while
NdMgAl$_{11}$O$_{19}$ exhibits strong single-ion anisotropy and frustration that
suppress long-range order down to 45~mK~\cite{Kumar2025NdMgAl11O19}. The
non-Kramers compound PrMgAl$_{11}$O$_{19}$, by contrast, shows a split
quasi-doublet and induced quantum-Ising-like dynamics with much larger moment
and interaction scales~\cite{Kumar2025}. In this landscape, SmMgAl$_{11}$O$_{19}$
stands out as a strongly quenched, nearly single-ion reference point, in which
the ground-state doublet is renormalized by both crystal-field effects and
$J$-multiplet mixing, complementary to the exchange-dominated QSL-like regime
exemplified by SmTa$_7$O$_{19}$ and to the more strongly interacting Kramers and
non-Kramers hexaaluminates.

\section{Conclusions}

In summary, SmMgAl$_{11}$O$_{19}$ realizes a Sm$^{3+}$ Kramers
doublet on a triangular lattice with strongly quenched effective moment,
in the extreme weak-exchange limit. Magnetometry and $c$-axis specific heat
consistently point to a low-field correlated regime in which part of the doublet
entropy remains locked in weakly interacting or disordered degrees of freedom,
and the field response deviates from that of an ideal two-level system. Under
applied field along $c$, the system crosses over into an almost single-ion
regime governed by a Zeeman-split Kramers doublet with a strongly reduced
effective $g$ factor ($g_c \simeq 0.6$) and a nearly recovered doublet entropy.
The modest anisotropy of the bulk response is best understood as arising from
anisotropic matrix elements and Van~Vleck contributions associated with
crystal-field splitting and $J$-multiplet mixing, rather than from any large
exchange anisotropy.

Looking ahead, local-probe $\mu$SR down to dilution temperatures can test for
ultra-low-$T$ freezing or persistent slow dynamics and quantify internal-field
distributions in the correlated regime. Extending specific heat and
magnetization measurements into the dilution-refrigerator range will further
constrain the fate of the residual entropy and the nature of the low-field
correlations. More broadly, SmMgAl$_{11}$O$_{19}$ provides a nearly single-ion,
strongly quenched reference point within the LnMgAl$_{11}$O$_{19}$ family,
complementary to the more strongly interacting Kramers and non-Kramers members
and to exchange-dominated Sm-based triangular magnets such as SmTa$_7$O$_{19}$.

\begin{acknowledgments}
We acknowledge funding from Charles University in Prague within the Grant Agency of Univerzita Karlova (grant number 438425). The work was also supported by the Ministry of Education, Youth and Sports of the Czech Republic through program INTER-EXCELLENCE II INTER-ACTION (LUABA24056). Crystal growth, structural analysis, and magnetic properties measurements were carried out in the MGML (\url{http://mgml.eu/}), supported within the Czech Research Infrastructures program (project no. LM2023065). This work was further supported by the Grant Agency of the Czech Republic (grant number 26-23051S).

\end{acknowledgments}

\bibliographystyle{aipnum4-2}

\bibliography{main}

@article{Ramirez2025,
  author = {Ramirez, A. P. and Syzranov, S. V.},
  title = {Short-range order and hidden energy scale in geometrically frustrated magnets},
  journal = {Mater. Adv.},
  volume = {6},
  pages = {1213},
  year = {2025},
  doi = {10.1039/D4MA00914B}
}

@article{Balents2010,
  author = {Balents, L.},
  title = {Spin liquids in frustrated magnets},
  journal = {Nature},
  volume = {464},
  pages = {199},
  year = {2010},
  doi = {10.1038/nature08917}
}

@article{Savary2017,
  author = {Savary, L. and Balents, L.},
  title = {Quantum spin liquids: a review},
  journal = {Rep. Prog. Phys.},
  volume = {80},
  pages = {016502},
  year = {2017},
  doi = {10.1088/0034-4885/80/1/016502}
}

@article{Broholm2020,
  author = {Broholm, C. and Cava, R. J. and Kivelson, S. A. and Nocera, D. G. and Norman, M. R. and Senthil, T.},
  title = {Quantum spin liquids},
  journal = {Science},
  volume = {367},
  number = {6475},
  pages = {eaay0668},
  year = {2020},
  doi = {10.1126/science.aay0668}
}

@article{Li2015,
  author = {Li, Yuesheng and Liao, Haijun and Zhang, Zhen and Li, Shiyan and Jin, Feng and Ling, Langsheng and Zhang, Lei and Zou, Youming and Pi, Li and Yang, Zhaorong and Wang, Junfeng and Wu, Zhonghua and Zhang, Qingming},
  title = {Gapless quantum spin liquid ground state in the two-dimensional spin-1/2 triangular antiferromagnet YbMgGaO$_4$},
  journal = {Sci. Rep.},
  volume = {5},
  pages = {16419},
  year = {2015},
  doi = {10.1038/srep16419}
}

@article{Rao2021,
  author = {Rao, X. and Hussain, G. and Huang, Q. and Chu, W. J. and Li, N. and Zhao, X. and Dun, Z. and Choi, E. S. and Asaba, T. and Chen, L. and Li, L. and Yue, X. Y. and Wang, N. N. and Cheng, J.-G. and Gao, Y. H. and Shen, Y. and Zhao, J. and Chen, G. and Zhou, H. D. and Sun, X. F.},
  title = {Survival of itinerant excitations and quantum spin state transitions in YbMgGaO$_4$ with chemical disorder},
  journal = {Nat. Commun.},
  volume = {12},
  pages = {4949},
  year = {2021},
  doi = {10.1038/s41467-021-25247-6}
}

@article{Shen2018YbMgGaO4,
  author = {Shen, Y. and Li, Y.-D. and Walker, H. C. and Steffens, P. and Boehm, M. and Zhang, X. and Shen, S. and Wo, H. and Chen, G. and Zhao, J.},
  title = {Fractionalized excitations in the partially magnetized spin liquid candidate {YbMgGaO4}},
  journal = {Nat. Commun.},
  volume = {9},
  pages = {4138},
  year = {2018},
  doi = {10.1038/s41467-018-06588-1}
}

@article{Li2015GaplessQSL,
  author = {Li, Y. and Liao, H. and Zhang, Z. and Li, S. and Jin, F. and Ling, L. and Zhang, L. and Zou, Y. and Pi, L. and Yang, Z. and Wang, J. and Wu, Z. and Zhang, Q.},
  title = {Gapless quantum spin liquid ground state in the two-dimensional spin-1/2 triangular antiferromagnet {YbMgGaO4}},
  journal = {Sci. Rep.},
  volume = {5},
  pages = {16419},
  year = {2015},
  doi = {10.1038/srep16419}
}

@article{Zhu2017YbMgGaO4,
  author = {Zhu, Z. and Maksimov, P. A. and White, S. R. and Chernyshev, A. L.},
  title = {Disorder-Induced Mimicry of a Spin Liquid in {YbMgGaO4}},
  journal = {Phys. Rev. Lett.},
  volume = {119},
  pages = {157201},
  year = {2017},
  doi = {10.1103/PhysRevLett.119.157201}
}

@article{Li2017CEF,
  author = {Li, Y. and Adroja, D. and Bewley, R. I. and Voneshen, D. and Tsirlin, A. A. and Gegenwart, P. and Zhang, Q.},
  title = {Crystalline Electric-Field Randomness in the Triangular Lattice Spin-Liquid {YbMgGaO4}},
  journal = {Phys. Rev. Lett.},
  volume = {118},
  pages = {107202},
  year = {2017},
  doi = {10.1103/PhysRevLett.118.107202}
}

@article{Li2016MuonYbMgGaO4,
  author = {Li, Y. and Adroja, D. and Biswas, P. K. and Baker, P. J. and Zhang, Q. and Liu, J. and Tsirlin, A. A. and Gegenwart, P. and Zhang, Q.},
  title = {Muon Spin Relaxation Evidence for the {U(1)} Quantum Spin-Liquid Ground State in the Triangular Antiferromagnet {YbMgGaO4}},
  journal = {Phys. Rev. Lett.},
  volume = {117},
  pages = {097201},
  year = {2016},
  doi = {10.1103/PhysRevLett.117.097201}
}

@article{Norman2016Herbertsmithite,
  author = {Norman, M. R.},
  title = {Colloquium: Herbertsmithite and the search for the quantum spin liquid},
  journal = {Rev. Mod. Phys.},
  volume = {88},
  pages = {041002},
  year = {2016},
  doi = {10.1103/RevModPhys.88.041002}
}

@article{Huang2021Herbertsmithite,
  author = {Huang, Y. Y. and Xu, Y. and Wang, L. and Zhao, C. C. and Tu, C. P. and Ni, J. M. and Wang, L. S. and Pan, B. L. and Fu, Y.},
  title = {Heat Transport in Herbertsmithite: Can a Quantum Spin Liquid Survive Disorder?},
  journal = {Phys. Rev. Lett.},
  volume = {127},
  pages = {267202},
  year = {2021},
  doi = {10.1103/PhysRevLett.127.267202}
}

@article{Smaha2020Barlowite,
  author = {Smaha, R. W. and He, W. and Jiang, J. M. and Wen, J. and Jiang, Y.-F. and Sheckelton, J. P. and Titus, C. J. and Wang, S. G. and Chen, Y.-S. and Teat, S. J. and Aczel, A. A. and Zhao, Y. and Xu, G. and Lynn, J. W. and Jiang, H.-C. and Lee, Y. S.},
  title = {Materializing rival ground states in the barlowite family of kagome magnets: quantum spin liquid, spin ordered, and valence bond crystal states},
  journal = {npj Quantum Mater.},
  volume = {5},
  pages = {23},
  year = {2020},
  doi = {10.1038/s41535-020-0222-8}
}

@article{Tustain2020ZnBarlowite,
  author = {Tustain, K. and Ward-O'Brien, B. and Bert, F. and Han, T. and Luetkens, H. and Lancaster, T. and Huddart, B. M. and Baker, P. J. and Clark, L.},
  title = {From magnetic order to quantum disorder in the {Zn}-barlowite series of {S = 1/2} kagom{\'e} antiferromagnets},
  journal = {npj Quantum Mater.},
  volume = {5},
  pages = {74},
  year = {2020},
  doi = {10.1038/s41535-020-00276-4}
}

@article{Shaginyan2013Herbertsmithite,
  author = {Shaginyan, V. R. and Popov, K. G. and Khodel, V. A.},
  title = {Strongly correlated quantum spin liquid in herbertsmithite},
  journal = {J. Exp. Theor. Phys.},
  volume = {116},
  pages = {848--853},
  year = {2013},
  doi = {10.1134/S1063776113050245}
}

@article{Bramwell2001SpinIce,
  author = {Bramwell, S. T. and Gingras, M. J. P.},
  title = {Spin Ice State in Frustrated Magnetic Pyrochlore Materials},
  journal = {Science},
  volume = {294},
  number = {5546},
  pages = {1495--1501},
  year = {2001},
  doi = {10.1126/science.1064761}
}

@article{Gingras2014QSI,
  author = {Gingras, M. J. P. and McClarty, P. A.},
  title = {Quantum spin ice: a search for gapless quantum spin liquids in pyrochlore magnets},
  journal = {Rep. Prog. Phys.},
  volume = {77},
  pages = {056501},
  year = {2014},
  doi = {10.1088/0034-4885/77/5/056501}
}

@article{Greedan2006Pyrochlores,
  author = {Greedan, J. E.},
  title = {Frustrated rare earth magnetism: Spin glasses, spin liquids and spin ices in pyrochlore oxides},
  journal = {J. Alloys Compd.},
  volume = {408--412},
  pages = {444--455},
  year = {2006},
  doi = {10.1016/j.jallcom.2004.12.084}
}

@article{Savary2016BreathingPyrochlore,
  author = {Savary, L. and Wang, X. and Kee, H.-Y. and Kim, Y. B. and Yu, Y. and Chen, G.},
  title = {Quantum spin ice on the breathing pyrochlore lattice},
  journal = {Phys. Rev. B},
  volume = {94},
  pages = {075146},
  year = {2016},
  doi = {10.1103/PhysRevB.94.075146}
}

@article{Huang2016SpinIceHeisenberg,
  author = {Huang, Y. and Chen, K. and Deng, Y. and Prokof'ev, N. and Svistunov, B.},
  title = {Spin-Ice State of the Quantum Heisenberg Antiferromagnet on the Pyrochlore Lattice},
  journal = {Phys. Rev. Lett.},
  volume = {116},
  pages = {177203},
  year = {2016},
  doi = {10.1103/PhysRevLett.116.177203}
}

@article{Yao2020PyrochloreU1,
  author = {Yao, X.-P. and Li, Y.-D. and Chen, G.},
  title = {Pyrochlore {U(1)} spin liquid of mixed-symmetry enrichments in magnetic fields},
  journal = {Phys. Rev. Research},
  volume = {2},
  pages = {013334},
  year = {2020},
  doi = {10.1103/PhysRevResearch.2.013334}
}

@article{An2025NonKramersPyrochlore,
  author = {An, T. and Desrochers, F. and Kim, Y. B.},
  title = {Quantum spin liquids in pyrochlore magnets with non-Kramers local moments},
  journal = {Phys. Rev. B},
  volume = {112},
  pages = {195109},
  year = {2025},
  doi = {10.1103/g5zf-hrb5}
}

@article{Desrochers2022Ce2Zr2O7,
  author = {Desrochers, F. and Chern, L. E. and Kim, Y. B.},
  title = {Competing {U(1)} and {Z\_2} dipolar-octupolar quantum spin liquids on the pyrochlore lattice: Application to {Ce\_2Zr\_2O\_7}},
  journal = {Phys. Rev. B},
  volume = {105},
  pages = {035149},
  year = {2022},
  doi = {10.1103/PhysRevB.105.035149}
}

@article{Yahne2024Ce2Sn2O7,
  author = {Yahne, D. R. and Placke, B. and Sch{\"a}fer, R. and Benton, O. and Moessner, R. and Powell, M. and Kolis, J. W. and Pasco, C. M. and May, A. F. and Frontzek, M. D. and Smith, E. M. and Gaulin, B. D. and Calder, S. and Ross, K. A.},
  title = {Dipolar Spin Ice Regime Proximate to an All-In-All-Out N{\'e}el Ground State in the Dipolar-Octupolar Pyrochlore {Ce\_2Sn\_2O\_7}},
  journal = {Phys. Rev. X},
  volume = {14},
  pages = {011005},
  year = {2024},
  doi = {10.1103/PhysRevX.14.011005}
}

@article{Saber1981,
  author = {Saber, D. and Lejus, A. M.},
  title = {Elaboration and characterization of lanthanide aluminate single crystals with the formula LnMgAl$_{11}$O$_{19}$},
  journal = {Mater. Res. Bull.},
  volume = {16},
  number = {10},
  pages = {1325--1330},
  year = {1981},
  doi = {10.1016/0025-5408(81)90104-5}
}

@article{Gasperin1984,
  author = {Gasperin, M. and Saine, M. C. and Kahn, A. and Laville, F. and Lejus, A. M.},
  title = {Influence of M$^{2+}$ ions substitution on the structure of lanthanum hexaaluminates with magnetoplumbite structure},
  journal = {J. Solid State Chem.},
  volume = {54},
  number = {1},
  pages = {61--69},
  year = {1984},
  doi = {10.1016/0022-4596(84)90131-2}
}

@article{Kumar2025,
  author = {Kumar, S. and Klicpera, M. and Eliáš, A. and Kratochvílová, M. and Kancko, A. and Correa, C. and Załęski, K. and Śliwińska-Bartkowiak, M. and Colman, R. H. and Bastien, G.},
  title = {Induced quantum magnetism on a triangular lattice of non-Kramers ions in PrMgAl$_{11}$O$_{19}$},
  journal = {Phys. Rev. B},
  volume = {111},
  pages = {174444},
  year = {2025},
  doi = {10.1103/PhysRevB.111.174444}
}

@article{Kumar2025NdMgAl11O19,
  author = {Kumar, S. and Prokle{\v{s}}ka, J. and Za{\l}{\k{e}}ski, K. and Kancko, A. and Correa, C. A. and Sliwi{\'n}ska-Bartkowiak, M. and Bastien, G. and Colman, R. H.},
  title = {Magnetic Anisotropy and Absence of Long-Range Order in the Triangular Magnet {NdMgAl\_{11}O\_{19}}},
  journal = {arXiv e-prints},
  year = {2025},
  eid = {arXiv:2505.18898},
  url = {https://arxiv.org/abs/2505.18898},
  note = {arXiv:2505.18898 [cond-mat.str-el]}
}

@article{Bastien2025,
  author = {Bastien, G. and Eliáš, A. and Anderle, V. and Kancko, A. and Corrêa, C. A. and Kumar, S. and Proschek, P. and Prokleška, J. and Nádherný, L. and Sedmidubský, D. and Treu, T. and Gegenwart, P. and Kratochvílová, M. and Žonda, M. and Colman, R. H.},
  title = {Quantum disordered ground state and relative proximity to an exactly solvable model in the frustrated magnet CeMgAl$_{11}$O$_{19}$},
  journal = {arXiv preprint arXiv:2506.16207},
  year = {2025}
}

@article{Cao2024,
  author = {Cao, Yantao and Bu, Huanpeng and Fu, Zhendong and Zhao, Jinkui and Gardner, Jason S. and Ouyang, Zhongwen and Tian, Zhaoming and Li, Zhiwei and Guo, Hanjie},
  title = {Synthesis, disorder and Ising anisotropy in a new spin liquid candidate PrMgAl$_{11}$O$_{19}$},
  journal = {Mater. Futures},
  volume = {3},
  number = {3},
  pages = {035201},
  year = {2024},
  doi = {10.1088/2752-5724/ad4a93}
}

@article{Li2024PrMgAl11O19,
  author = {Li, N. and Rutherford, A. and Wang, Y. Y. and Liang, H. and Li, Q. J. and Zhang, Z. J. and Wang, H. and Xie, W. and Zhou, H. D.},
  title = {Ising-type quantum spin liquid state in {PrMgAl\_{11}O\_{19}}},
  journal = {Phys. Rev. B},
  volume = {110},
  pages = {134401},
  year = {2024},
  doi = {10.1103/PhysRevB.110.134401}
}

@article{Ma2024PrMgAl11O19,
  author = {Ma, Z. and Zheng, S. and Chen, Y. and Xu, R. and Dong, Z.-Y. and Wang, J. and Du, H. and Embs, J. P. and Li, S. and Li, Y. and Zhang, Y. and Liu, M. and Zhong, R. and Liu, J.-M. and Wen, J.},
  title = {Possible gapless quantum spin liquid behavior in the triangular-lattice Ising antiferromagnet {PrMgAl\_{11}O\_{19}}},
  journal = {Phys. Rev. B},
  volume = {109},
  pages = {165143},
  year = {2024},
  doi = {10.1103/PhysRevB.109.165143}
}

@article{Tu2024PrMgAl11O19,
  author = {Tu, C. P. and Ma, Z. and Wang, H. R. and Jiao, Y. H. and Dai, D. Z. and Li, S. Y.},
  title = {Gapped quantum spin liquid in a triangular-lattice Ising-type antiferromagnet {PrMgAl\_{11}O\_{19}}},
  journal = {Phys. Rev. Research},
  volume = {6},
  pages = {043147},
  year = {2024},
  doi = {10.1103/PhysRevResearch.6.043147}
}

@article{Cao2025CeMgAl11O19,
  author = {Cao, Y. and Koda, A. and Le, M. D. and Pomjakushin, V. and Liu, B. and Fu, Z. and Li, Z. and Zhao, J. and Tian, Z. and Guo, H.},
  title = {U(1) Dirac quantum spin liquid candidate in triangular-lattice antiferromagnet {CeMgAl\_{11}O\_{19}}},
  journal = {Sci. China Phys. Mech. Astron.},
  volume = {68},
  pages = {267011},
  year = {2025},
  doi = {10.1007/s11433-024-2448-1}
}

@article{Gao2024CeMgAl11O19,
  author = {Gao, B. and Chen, T. and Liu, C. and Klemm, M. L. and Zhang, S. and Ma, Z. and Xu, X. and Won, C. and McCandless, G. T. and Murai, N. and Ohira-Kawamura, S. and Moxim, S. J. and Ryan, J. T. and Huang, X. and Wang, X. and Chan, J. Y. and Cheong, S.-W. and Tchernyshyov, O. and Balents, L. and Dai, P.},
  title = {Spin Excitation Continuum in the Exactly Solvable Triangular-Lattice Spin Liquid {CeMgAl\_{11}O\_{19}}},
  journal = {arXiv e-prints},
  year = {2024},
  eid = {arXiv:2408.15957},
  doi = {10.48550/arXiv.2408.15957},
  url = {https://arxiv.org/abs/2408.15957},
  note = {arXiv:2408.15957 [cond-mat.str-el]}
}

@article{Sanders2017KBaREBO3,
  author  = {Sanders, M. B. and Cevallos, F. A. and Cava, R. J.},
  title   = {Magnetism in the KBaRE(BO$_3$)$_2$ (RE = Sm, Eu, Gd, Tb, Dy, Ho, Er, Tm, Yb, Lu) series: materials with a triangular rare earth lattice},
  journal = {Mater. Res. Express},
  volume  = {4},
  pages   = {036102},
  year    = {2017},
  doi     = {10.1088/2053-1591/aa60a2}
}

@article{Sanders2016RE3Sb3Mg2O14,
  author  = {Sanders, M. B. and Baroudi, K. M. and Krizan, J. W. and Mukadam, O. A. and Cava, R. J.},
  title   = {Synthesis, crystal structure, and magnetic properties of RE$_3$Sb$_3$Mg$_2$O$_{14}$ (RE = La, Pr, Sm, Eu, Tb, Ho): new 2D Kagome materials},
  journal = {arXiv preprint arXiv:1601.06639},
  year    = {2016},
  doi     = {10.48550/arXiv.1601.06639}
}

@article{Bairwa2025SmTa7O19,
  author = {Bairwa, D. and Bandyopadhyay, A. and Adroja, D. and Stenning, G. B. G. and Luetkens, H. and Hicken, T. J. and Krieger, J. A. and Cibin, G. and Rotter, M. and Rayaprol, S. and Babu, P. D. and Elizabeth, S.},
  title = {Quantum spin liquid ground state in the rare-earth triangular antiferromagnet {SmTa\_7O\_{19}}},
  journal = {Phys. Rev. B},
  volume = {111},
  pages = {104413},
  year = {2025},
  doi = {10.1103/PhysRevB.111.104413}
}

@article{Arh2022,
  author = {Arh, T. and Sana, B. and Pregelj, M. and Khuntia, P. and Jagličić, Z. and Le, M. Duc and Biswas, P. K. and Manuel, P. and Mangin-Thro, L. and Ozarowski, A. and Zorko, A.},
  title = {The Ising triangular-lattice antiferromagnet neodymium heptatantalate as a quantum spin liquid candidate},
  journal = {Nat. Mater.},
  volume = {21},
  pages = {416--422},
  year = {2022},
  doi = {10.1038/s41563-021-01169-y}
}

@misc{CrysAlisPro2024,
  author = {{Rigaku Oxford Diffraction}},
  title = {CrysAlisPro Software, Version 1.171.43.143a},
  year = {2024},
  note = {Rigaku Oxford Diffraction}
}

@article{Clark1995,
  author = {Clark, R. C. and Reid, J. S.},
  title = {The analytical calculation of absorption in multifaceted crystals},
  journal = {Acta Crystallogr. Sect. A},
  volume = {51},
  pages = {887--897},
  year = {1995},
  doi = {10.1107/S0108767395007367}
}

@article{Petricek2023,
  author = {Petříček, Václav and Palatinus, Lukáš and Plášil, Jakub and Dušek, Michal},
  title = {JANA2020 – a new version of the crystallographic computing system JANA},
  journal = {Z. Kristallogr. Cryst. Mater.},
  volume = {238},
  pages = {271},
  year = {2023},
  doi = {10.1515/zkri-2023-0005}
}

@article{Palatinus2007,
  author = {Palatinus, L. and Chapuis, G.},
  title = {SUPERFLIP -- a computer program for the solution of crystal structures by charge flipping in arbitrary dimensions},
  journal = {J. Appl. Cryst.},
  volume = {40},
  pages = {786--790},
  year = {2007},
  doi = {10.1107/S0021889807029238}
}

@article{Scheie2021,
  author = {Scheie, A.},
  title = {PyCrystalField: software for calculation, analysis and fitting of crystal electric field Hamiltonians},
  journal = {J. Appl. Cryst.},
  volume = {54},
  pages = {356--362},
  year = {2021},
  doi = {10.1107/S160057672001554X}
}

@article{PhysRevB.99.134415,
  title = {Intermultiplet transitions and magnetic long-range order in Sm-based pyrochlores},
  author = {Pe\ifmmode \mbox{\c{c}}\else \c{c}\fi{}anha-Antonio, Viviane and Feng, Erxi and Sun, Xiao and Adroja, Devashibhai and Walker, Helen C. and Gibbs, Alexandra S. and Orlandi, Fabio and Su, Yixi and Br\"uckel, Thomas},
  journal = {Phys. Rev. B},
  volume = {99},
  issue = {13},
  pages = {134415},
  numpages = {12},
  year = {2019},
  month = {Apr},
  publisher = {American Physical Society},
  doi = {10.1103/PhysRevB.99.134415},
  url = {https://link.aps.org/doi/10.1103/PhysRevB.99.134415}
}

@book{AbragamBleaney1970,
  author    = {Abragam, A. and Bleaney, B.},
  title     = {Electron Paramagnetic Resonance of Transition Ions},
  publisher = {Oxford University Press},
  address   = {Oxford},
  year      = {1970}
}

@article{PecanhaAntonio2019_SmPyrochlores,
  title     = {Intermultiplet transitions and magnetic long-range order in Sm-based pyrochlores},
  author    = {Pe{\c{c}}anha-Antonio, Viviane and Feng, Erxi and Sun, Xiao and Adroja, Devashibhai and Walker, Helen C. and Gibbs, Alexandra S. and Orlandi, Fabio and Su, Yixi and Br{\"u}ckel, Thomas},
  journal   = {Physical Review B},
  volume    = {99},
  number    = {13},
  pages     = {134415},
  year      = {2019},
  publisher = {American Physical Society},
  doi       = {10.1103/PhysRevB.99.134415},
  url       = {https://doi.org/10.1103/PhysRevB.99.134415}
}

@book{Wagner1972_TheoryOfMagnetism,
  title     = {Introduction to the Theory of Magnetism},
  author    = {Wagner, D.},
  series    = {International Series in Natural Philosophy},
  volume    = {48},
  year      = {1972},
  publisher = {Pergamon Press},
  address   = {Oxford}
}

\end{document}